\documentclass[journal,comsoc]{IEEEtran}
\IEEEoverridecommandlockouts

\usepackage{bbm}
\usepackage{graphicx}
\usepackage{amssymb}
\usepackage{amsmath}
\usepackage{cite}
\usepackage{mathrsfs}
\usepackage[displaymath,mathlines]{lineno}
\usepackage{color}
\usepackage{overpic}
\usepackage{multirow}
\usepackage{algpseudocode}
\usepackage{algorithm,algpseudocode}
\usepackage{algorithmicx}
\usepackage{pbox}
\usepackage{multicol}
\usepackage{lipsum}
\usepackage{amsthm}
\usepackage{lipsum}
\usepackage{epstopdf}
\usepackage{makecell}
\usepackage{diagbox}
\usepackage{enumitem}
\usepackage{xcolor}

\makeatletter
\newcommand{\doublewidetilde}[1]{{%
		\mathpalette\double@widetilde{#1}%
}}
\newcommand{\double@widetilde}[2]{%
	\sbox\z@{$\m@th#1\widetilde{#2}$}%
	\ht\z@=.5\ht\z@
	\widetilde{\box\z@}%
}
\makeatother

\newtheorem{theorem}{Theorem}
\newtheorem{lemma}{Lemma}

\newtheorem{remark}{Remark}

\begin{document}

\title{Joint Power Allocation and Load Balancing Optimization for Energy-Efficient Cell-Free Massive MIMO Networks}
%\title{}

\author{\IEEEauthorblockN{Trinh Van Chien, Emil Bj\"{o}rnson, \textit{Senior Member}, \textit{IEEE},  and Erik G. Larsson, \textit{Fellow}, \textit{IEEE}}
	\thanks{
		T. V. Chien was with the Department of Electrical
		Engineering (ISY), Link\"{o}ping University, 581 83 Link\"{o}ping,
		Sweden. He is now with the School of Electronics and Telecommunications, Hanoi University of Science and Technology, Hanoi, Vietnam (email: trinhchien.dt3@gmail.com). E. Bj\"{o}rnson and E. G. Larsson are with the Department of Electrical
		Engineering (ISY), Link\"{o}ping University, 581 83 Link\"{o}ping,
		Sweden (email: emil.bjornson@liu.se; erik.g.larsson@liu.se). This paper was supported by ELLIIT and CENIIT. Parts of this paper were presented at ICASSP 2020 \cite{Chien2020}.
	}
}

\maketitle

\begin{abstract}
Large-scale distributed antenna systems with many access points (APs) that serve the users by coherent joint transmission is being considered for 5G-and-beyond networks. The technology is called Cell-free Massive MIMO and can provide a more uniform service level to the users than a conventional cellular topology.
For a given user set, only a subset of the APs is likely needed to satisfy the users' performance demands, particularly outside the peak traffic hours. To find achieve an energy-efficient load balancing, we minimize the total downlink power consumption at the APs, considering both the transmit powers and hardware dissipation. APs can be temporarily turned off to reduce the latter part. The formulated optimization problem is non-convex but, nevertheless, a globally optimal solution is obtained by solving a mixed-integer second-order cone program. Since the computational complexity is prohibitive for real-time implementation, we also propose two low-complexity algorithms that exploit the inherent group-sparsity and the optimized transmit powers in the problem formulation. Numerical results manifest that our optimization algorithms can greatly reduce the power consumption compared to keeping all APs turned on and only minimizing the transmit powers. Moreover, the low-complexity algorithms can effectively handle the power allocation and AP activation for large-scale networks.
\end{abstract}

\begin{IEEEkeywords}
Cell-free Massive MIMO, sparse optimization, total power minimization, energy efficiency.
\end{IEEEkeywords}% no keywords

\IEEEpeerreviewmaketitle

\section{Introduction}

The use of mobile phones and other portable devices is continuously increasing the demand for data in wireless networks \cite{Andrews2014a, tariq2019speculative}. The cellular technology has evolved over time to cater for the increasing demand but although Massive MIMO (multiple-input multiple-output) is now being used, beamforming can only mitigate the large pathloss variations in cellular deployments; cell-edge users might have a 50\,dB weaker channel than the cell-center users, and Massive MIMO with 100 antenna can only compensate for 20\,dB of that. Cell-free Massive MIMO is a promising new technology to deal with the mediocre cell-edge performance by distributing the antennas over the coverage area and removing the cell edges by joint operation \cite{Ngo2017a, Nayebi2017a, nguyen2018optimal}. Each distributed antenna location is called an access point (AP) and the APs transmit coherently in the downlink and process their received signals coherently in the uplink, leading to a higher signal-to-noise ratio (SNR) without using more transmit power.
 The coherent joint transmission in Cell-free Massive MIMO is inherited from the classical coordinated multipoint (CoMP) beamforming design with a few co-located antenna arrays \cite{Shamai2001a, Tolli2008a, Bjornson2010c, Boldi2011a,Peng2014a} and gradually extended to scenarios with many distributed APs with few antennas. These previous network designs were mainly considering slowly fading channels where the  small-scale fading realizations can be estimated perfectly and spectral efficiency (SE) utility metrics can be formulated as functions of one set of small-scale fading realizations.
 In contrast, a key novelty in the Cell-free Massive MIMO area is the analysis of practical fast fading channels, for which the ergodic SE is the preferred performance metric and the SE depends on imperfect channel state information (CSI) and pilot contamination. In scenarios with Rayleigh fading and some choices of linear processing, the ergodic SE of Cell-free Massive MIMO can even be obtained in closed form, which makes it easier to formulate and solve practical spatial resource allocation problems.

There will be $12.3$ billion wirelessly connected devices by $2022$ \cite{index2019global}, which raises concerns about the power consumption and the corresponding energy-related pollution. Cellular networks have been developed to maximize the SE and coverage, leading to the norm of transmitting at the maximum allowed power in the downlink \cite{Auer2011a}. This results in high power consumption at the base stations/APs, even when the traffic is low. Upcoming technologies should be redesigned to achieve a direct connection between power consumption and traffic load \cite{frenger2011reducing}, so that the power is low when the users request low SEs.
In the context of Cell-free Massive MIMO, energy efficiency optimization has been considered in \cite{ngo2018total, nguyen2017energy}. These papers considered how the fronthaul power consumption can be reduced by only serving each user by a subset of the APs, but all APs are assumed to be turned on continuously. As reported in \cite{feng2017boost, vu2015energy, han2016survey} (and references therein), the energy efficiency of heterogeneous or cloud radio access networks can be substantially improved by also turning APs on and off. However, the operating point where the energy efficiency is maximized might not provide the service quality that the users need. Hence, the goal of \emph{load balancing} is to map the current traffic load to the available transmission resources of the network in a more efficient fashion. The authors \cite{luo2014downlink} instead considered that each user has an SE requirement that the system must satisfy with minimum power consumption, considering both the transmit power and hardware-consumed power of active APs. Hence, the goal of the resource allocation under load balancing is for the system to deliver the required SEs with as low total power consumption as possible. These previous works considered cellular networks with deterministic (or slowly fading) channels and perfect CSI, where the channel takes one realization throughout the entire transmission, and therefore the optimization problems are formulated based on one channel realization. This modeling is only appropriate in special cases where the users are entirely static. In contrast, the Cell-free Massive MIMO methodology enables analysis of realistic fast fading channels, where the ergodic SE is the appropriate performance metric and CSI imperfections (including pilot contamination) are unavoidable. 
To the best of our knowledge, there is no previous work on AP activation in Cell-free Massive MIMO networks. 

\subsection{Main Contributions}
Motivated by the coexistence of multiple users using different services with stringent requirements, this paper considers that each user has a predetermined downlink ergodic SE requirement that the network must satisfy to avoid interrupting any of the users' services. The users and APs are arbitrarily distributed, thus it is likely that these SE requirements can be fulfilled without using all the APs. When minimizing the total power consumption in the downlink, we consider both the transmit power and the hardware-consumed power. Bearing in mind that each user will mainly be served by its neighboring APs, we consider the possibility to turn off APs that are not needed to serve the current set of users. This is an important feature since Cell-Free Massive MIMO networks may have many APs \cite{Ngo2017a, Nayebi2017a}, where the large number is needed to provide consistent coverage but might not be needed at every time instant. We formulate the new optimization problem using rigorous closed-form ergodic SE expressions for uncorrelated Rayleigh fading channels, linear precoding (either maximum ratio transmission (MRT) or full-pilot zero-forcing (F-ZF)), imperfect CSI, and pilot contamination. This allows us to optimize large-scale networks with many APs and users. The main contributions are:
\begin{itemize}
	\item We formulate a total downlink power minimization problem, where the active APs and transmit power allocation are the optimization variables. This problem is non-convex, but we still can obtain a globally optimal solution to both the transmit power allocation and the active APs topology by solving a mixed-integer second order cone (SOC) program.	
	\item Since algorithms that solve mixed-integer SOC programs are too complex for real-time applications, two heuristic low-complexity algorithms are developed by exploiting the structure of the optimization problem. The first algorithm utilizes both the optimized transmit power and sparsity, while the second algorithm only utilizes optimized transmit powers to determine which APs to turn off.   
	\item Numerical results demonstrate that there are scenarios where only a subset of the APs are needed to satisfy the SE requirements for all users and large power reductions can be achieved by turning off the remaining APs. Moreover, the low-complexity algorithms give total power consumptions close to the global minimum.
\end{itemize}
The rest of this paper is organized as follows: Section~\ref{Sec:SysModel} gives the network model together with the downlink SE analysis. A power consumption model is introduced in Section~\ref{Sec:TotalPowerOpt}. Then, we formulate and solve the total power minimization problem to obtain the global optimum. Section~\ref{Sec:SubSols} propose two suboptimal algorithms with low complexity. Finally, Section~\ref{Sec:NumRes} presents extensive numerical results and the main conclusions are given in Section~\ref{Sec:Conclusion}.

\textit{Notations:} We use boldface lower-case and upper-case letters to denote vectors and matrices, respectively. The transpose is denoted by the superscript $(\cdot)^T$ and the Hermitian transpose is denoted by $(\cdot)^H$. The expectation operator is $\mathbb{E} \{ \cdot \}$ and $\mathcal{CN}(\cdot, \cdot)$ denotes a circularly symmetric complex Gaussian distribution. The Euclidean norm, $\ell_1$-norm, and $\ell_{p}$-norm of a vector $\mathbf{x}$ is denoted as $\| \mathbf{x} \|$, $\| \mathbf{x} \|_1$, and $\| \mathbf{x} \|_p$, respectively. Finally, the cardinality of the set $\mathcal{X}$ is denoted by $|\mathcal{X}|$ and $\mathcal{O} (\cdot)$ represents the big-$\mathcal{O}$ notation.

\section{System Model} \label{Sec:SysModel}
We consider a Cell-free Massive MIMO network with $M$ APs and $K$ users that are arbitrarily distributed over the coverage area. % as illustrated in Fig.~\ref{FigSystem}.
A central processing unit (CPU) is connected to all APs via unlimited fronthaul links. Each AP is equipped with $N$ antennas, while there is a single antenna in each user device. We assume every user has a required SE value [b/s/Hz] that must be satisfied. At a given time instance, the users will be heterogeneously distributed and their SE requirements are likely in the interior of the capacity region of the network. Intuitively, each user will receive most of its downlink signal power from the closest APs while more distant APs typically have a negligible impact. Hence, it might suffice to only utilize a subset $\mathcal{A} \subseteq \{1, \ldots M \}$ of the APs to satisfy the SE requirements. The remaining APs can be put into sleep mode to save power. The main goal of this paper is to find a subset $\mathcal{A}$ of active APs  and the corresponding transmit powers that satisfy the SE requirements while minimizing the total power consumption, taking the power dissipation in active APs into account.

Practical channels exhibit fast fading, which means that the channels vary rapidly over time and frequency during the communication.
We model this using the classic block fading model \cite{massivemimobook}, where the channel is fixed within a finite-sized time-frequency coherence interval and take independent random realizations in each such coherence interval.
A coherence interval encompasses $\tau_c$ symbols and $\tau_p$ of them are dedicated to estimate the channels from uplink pilot signals. We consider a time division duplex (TDD) protocol and focus on the downlink performance analysis, thus the remaining $\tau_c - \tau_p$ symbols are used for downlink data transmission. The channel response between AP~$m$ and user~$k$ is denoted by $\mathbf{h}_{mk} \in \mathbb{C}^N$ and is assumed to follow an independent and identically distributed Rayleigh fading model:
\begin{equation}
\mathbf{h}_{mk} \sim \mathcal{CN} (\mathbf{0}, \beta_{mk} \mathbf{I}_N),
\end{equation}
where $\beta_{mk} \geq 0$ denotes the large-scale fading coefficient involving both path loss and shadowing. Each channel takes an independent realization in each coherence interval. We assume the APs know the channel statistics, but the realizations need to be estimated from the uplink pilots.

\subsection{Uplink Pilot Transmission}
In the uplink training phase, $\pmb{\Psi} = [\pmb{\psi}_1, \ldots, \pmb{\psi}_{\tau_p}] \in \mathbb{C}^{\tau_p \times \tau_p}$ is a matrix gathering a set of $\tau_p$ orthonormal pilot signals that are assigned to the $K$ users. Specifically, user~$k$ transmits the pilot signal $\sqrt{\tau_p} \pmb{\psi}_{i_k} \in \mathbb{C}^{\tau_p}$ with $i_k \in \{1, \ldots, \tau_p \}$ being the pilot index. We consider a fixed and arbitrary pilot assignment but note that many algorithms have been proposed in prior work \cite{interdonato2019ubiquitous,Bjornson2019Aug}. We let $\mathcal{P}_k$ denote the subset of users assigned to the same pilot signal as user~$k$, thus it holds that
\begin{equation}
\pmb{\psi}_{i_k}^H \pmb{\psi}_{i_{k'}} = \begin{cases}
1 & \mbox{ if } k' \in \mathcal{P}_k,\\
0 & \mbox{ if } k' \notin \mathcal{P}_k.
\end{cases}
\end{equation}
The signal $\mathbf{Y}_m \in \mathbb{C}^{N \times \tau_p}$ received at AP~$m$ is a superposition of the transmitted pilot signals from all the $K$ users:
\begin{equation}
\mathbf{Y}_m = \sum_{k=1}^{K} \sqrt{\tau_p p_k} \mathbf{h}_{mk} \pmb{\psi}_{i_k}^H + \mathbf{N}_m,
\end{equation}
where $p_k$ is the transmit pilot power of user~$k$ and $\mathbf{N}_m \in \mathbb{C}^{N \times \tau_p}$ is additive noise where each element is independently distributed as $\mathcal{CN}(0, \sigma_{\mathrm{UL}}^2)$. AP~$m$ computes an estimate of $\mathbf{h}_{mk}$ from the sufficient statistics $\mathbf{y}_{mk} = \mathbf{Y}_m \pmb{\psi}_{i_k} \in \mathbb{C}^{N}$, which is obtained as
\begin{equation} \label{eq:ymk}
\begin{split}
\mathbf{y}_{mk} =&  \sum_{k' \in \mathcal{P}_k} \sqrt{\tau_p p_{k'}} \mathbf{h}_{mk'} + \mathbf{N}_m \pmb{\psi}_{i_k}.
\end{split}
\end{equation}
\begin{lemma} \label{Lemma:EstChannel}
The minimum mean square error (MMSE) estimate of the channel between user~$k$ and AP~$m$ is
\begin{equation} \label{eq:EstChan}
\hat{\mathbf{h}}_{mk} =  \mathbb{E} \{\mathbf{h}_{mk}  | \mathbf{y}_{mk} \} = \frac{ \sqrt{\tau_p p_k } \beta_{mk} }{ \tau_p \sum_{k' \in \mathcal{P}_k}  p_{k'}\beta_{mk'} + \sigma_{\mathrm{UL}}^2 } \mathbf{y}_{mk}.
\end{equation}
The channel estimate is distributed as $\hat{\mathbf{h}}_{mk} \sim \mathcal{CN} \left( \mathbf{0}, \gamma_{mk} \mathbf{I}_N \right)$, in which the variance $\gamma_{mk}$ is 
\begin{equation}
\gamma_{mk} =  \frac{ \tau_p p_k  \beta_{mk}^2 }{ \tau_p \sum_{k' \in \mathcal{P}_k}  p_{k'}\beta_{mk'} + \sigma_{\mathrm{UL}}^2 }.
\end{equation}
\end{lemma}
\begin{IEEEproof}
The proof is adopted from the standard MMSE estimation \cite{Kay1993a} to our notation.
\end{IEEEproof}
Lemma~\ref{Lemma:EstChannel} gives the precise expression of the estimated channel between an arbitrary AP~$m$ and user~$k$. If AP~$m$ is in sleep mode, it will not estimate the channel and a convenient way to represent that is by substituting $\beta_{mk} =0$ into \eqref{eq:EstChan}, for all $k$, which leads to a zero-valued channel estimate.

\subsection{Downlink Performance Analysis} \label{Sec:DPL}
In the downlink data transmission phase, each active AP constructs the precoding vectors based on their locally estimated channels that were computed using Lemma~\ref{Lemma:EstChannel}. Let us denote the precoding vector used by AP~$m$ to steer the data signal to user~$k$ as $\mathbf{w}_{mk} \in \mathbb{C}^N$. Let $s_{k}$ denote the data symbol that is jointly transmitted to user~$k$ by all the active APs and assume $\mathbb{E} \{ |s_{k}|^2\} =1 $. The transmitted signal $\mathbf{x}_m \in \mathbb{C}^N$ at AP~$m$ to all $K$ users is
\begin{equation} \label{eq:xm}
\mathbf{x}_m = \sum_{k=1}^K \sqrt{\rho_{mk}} \mathbf{w}_{mk} s_{k},
\end{equation}
where $\rho_{mk} \geq 0$ is the transmit data power that AP~$m$ allocates to user~$k$.
The received signal at user~$k$ from all the active APs is
\begin{equation}
r_k = \sum_{m \in \mathcal{A}} \mathbf{h}_{mk}^H \mathbf{x}_m + \tilde{w}_k,
\end{equation}
where  $\tilde{w}_k \sim \mathcal{CN}(0, \sigma_{\mathrm{DL}}^2)$ is independent additive noise with the zero mean and the variance $\sigma_{\mathrm{DL}}^2$. 
By using the capacity bounding technique described in \cite[Section~2.3]{Marzetta2016a}, \cite[Section~4.3]{massivemimobook}, a lower bound on the ergodic channel capacity of user~$k$ is
\begin{multline} \label{eq:ErgRate1}
R_k = \left(1 - \frac{\tau_p}{\tau_c} \right) \times\\
\log_2 \left( 1 + \frac{|\mathsf{DS}_k|^2}{ \mathbb{E} \{ |\mathsf{BU}_k|^2 \} + \sum_{k' \neq k}^K \mathbb{E} \{ |\mathsf{UI}_{k'k}|^2 \} + \sigma_{\mathrm{DL}}^2 }   \right),
\end{multline}
where $\mathsf{DS}_k$, $\mathsf{BU}_k$, and $\mathsf{UI}_{k'k}$ terms denote the desired signal, the beamforming uncertainty gain, and the inter-user interference, respectively, which are expressed as
\begin{align}
&\mathsf{DS}_k =  \mathbb{E} \left\{\sum_{m \in \mathcal{A}} \sqrt{\rho_{mk}} \mathbf{h}_{mk}^H \mathbf{w}_{mk}\right\},\\
%\end{align}
%\begin{align}
&\mathsf{BU}_k  =  \sum_{m \in \mathcal{A}} \sqrt{\rho_{mk}} \mathbf{h}_{mk}^H \mathbf{w}_{mk} - \mathbb{E} \left\{ \sum_{m \in \mathcal{A}} \sqrt{\rho_{mk}} \mathbf{h}_{mk}^H \mathbf{w}_{mk} \right\},\\
%\end{align}
%\begin{align}
&\mathsf{UI}_{k'k} = \sum_{m \in \mathcal{A}} \sqrt{\rho_{mk'}} \mathbf{h}_{mk}^H \mathbf{w}_{mk'}.
\end{align}
We stress that the lower bound on the downlink channel capacity in \eqref{eq:ErgRate1} can be applied for any precoding scheme and any active AP set. To obtain closed-form expressions that can be efficiently used for optimization, we now assume the active APs either use MRT or F-ZF precoding, which are defined for $m \in \mathcal{A}$ as
\begin{equation} \label{eq:ModifiedMRT}
\mathbf{w}_{mk} = \begin{cases}
\frac{\hat{\mathbf{h}}_{mk}}{\sqrt{\mathbb{E} \{ \| \hat{\mathbf{h}}_{mk} \|^2 \}}} & \mbox{if MRT},\\
\frac{\widehat{\mathbf{H}}_{m} \left( \widehat{\mathbf{H}}_{m}^H \widehat{\mathbf{H}}_{m}  \right)^{-1} \mathbf{e}_{i_k}}{ \sqrt{ \mathbb{E} \left\{ \left\| \widehat{\mathbf{H}}_{m} \left( \widehat{\mathbf{H}}_{m}^H \widehat{\mathbf{H}}_{m}  \right)^{-1} \mathbf{e}_{i_k} \right\|^2 \right\}}} & \mbox{if F-ZF},
\end{cases}
\end{equation}
where $\widehat{\mathbf{H}}_{m} = \mathbf{Y}_{m} \pmb{\Psi} \in \mathbb{C}^{N \times K}$ and $\mathbf{e}_{i_k}$ is the $i_k$-th column of identity matrix $\mathbf{I}_{\tau_p}$. 
\begin{lemma} \label{lemma:ClosedFormMRC}
%In a Cell-free Massive MIMO network, 
The downlink ergodic SE of user~$k$ is
\begin{equation} \label{eq:DLRate}
R_k (\{ \rho_{mk} \}, \mathcal{A}) = \left( 1 - \frac{\tau_p}{\tau_c} \right)\log_2 \left(1 + \mathrm{SINR}_k (\{ \rho_{mk} \}, \mathcal{A}) \right),
\end{equation}
where the effective SINR is given in \eqref{eq:SINRk}.
\begin{figure*}
\begin{equation} \label{eq:SINRk}
%\fontsize{10}{10}
%{
\mathrm{SINR}_k (\{ \rho_{mk} \}, \mathcal{A}) = \frac{G \left( \sum_{ m \in \mathcal{A}} \sqrt{\rho_{mk} \gamma_{mk}}  \right)^2 }{ G \sum_{k' \in \mathcal{P}_k \setminus \{k\} } \left( \sum_{m \in \mathcal{A}} \sqrt{\rho_{mk'}\gamma_{mk}} \right)^2 + \sum_{k' =1}^K   \sum_{m \in \mathcal{A}} \rho_{mk'} z_{mk} +\sigma_{\mathrm{DL}}^2}.
%}
\end{equation}
\hrule
%\vspace*{-0.2cm}
\end{figure*}
The parameters $G$ and $z_{mk}$ depend on the selection of precoding scheme. MRT gives $G = N$ and $z_{mk} = \beta_{mk}$. For $N > \tau_p$, F-ZF gives $G= N- \tau_p$ and $z_{mk} = \beta_{mk} - \gamma_{mk}$.
\end{lemma}
\begin{IEEEproof}
The detailed proof aligns with \cite{Chien2016b} for MRT precoding and with \cite{interdonato2018} for F-ZF precoding, except for the different notation and that we only consider that a subset of the $M$ APs is in active mode. 
\end{IEEEproof}
In \eqref{eq:SINRk}, the numerator is proportional to $G$, which is the array gain from the multiple antennas installed at each AP. The fact that the contributions from different APs are summed up inside the square is typical for coherent joint transmission. The first part in the denominator represents coherent interference from other users in the set $\mathcal{P}_k$, which is caused by pilot contamination. The remaining parts are the non-coherent interference and noise. If F-ZF precoding is used, each AP ``sacrifices'' $\tau_p$ antennas (i.e., $\tau_p$ spatial degrees of freedom) to cancel interference between users that have different pilots. We stress that the condition on the number of antennas $N > \tau_p$ is essential for the validity of closed-form SE expression if F-ZF precoding is utilized.

The ergodic SE in \eqref{eq:DLRate} will hereafter be used to establish the SE constraint for each user in the network. Unlike the previous work \cite{Chien2016b, interdonato2018} that considered all APs in active mode $\mathcal{A} = \{1, \ldots, M \}$, the new closed-form SE expressions in \eqref{eq:DLRate} are multivariate functions of both the transmit powers and the set of active APs. One can observe that at least a single AP should be activated, say $1 \leq |\mathcal{A}| \leq M$, when the network serves $K \geq 1$ users with the non-zero SE requirements. We will use these expressions to formulate and solve a new total power consumption minimization problem for Cell-free Massive MIMO networks in the next sections. 

\section{Total Power Minimization Problem} \label{Sec:TotalPowerOpt}

To maximize the energy efficiency of the network, we can minimize the power consumption while satisfying the SE requirements of the users. This section formulates a new total power consumption minimization problem subject to transmit power constraints at the APs and the required SEs of the $K$ users. The optimization variables are the active AP set and the transmit powers. The global optimum can be found by an exhaustive search, but it is extremely costly, in particular for large networks, since the problem contains both the continuous transmit power variables and discrete variables representing the active APs. We reduce the computational complexity by transforming this non-convex problem into a mixed-integer SOC program, which is solved by the branch-and-bound approach.

\vspace{-4mm}

\subsection{Problem Formulation}
The power consumption of the network consists of both the transmit power and power dissipation in the transceiver hardware of the active APs.
%We first borrow the linear approximated power consumption model in
Similar to \cite{EARTH_D23_short, ngo2018total}, we model the total power consumption from the all active APs in the network as
\begin{equation} \label{eq:Ptotal}
\begin{split}
P_{\mathrm{total}} (\{ \rho_{mk} \}, \mathcal{A}) =& \sum_{m \in \mathcal{A}} \sum_{k=1}^K \Delta_m \rho_{mk} + \sum_{m \in \mathcal{A}} P_{m} \\
& + B \sum_{m \in \mathcal{A}} \sum_{k=1}^K P_{\mathrm{bt},m} R_k (\{ \rho_{mk} \}, \mathcal{A}), 
\end{split}
\end{equation}
where the first term in \eqref{eq:Ptotal} is the total transmit power consumed by every active AP. The transmit power at AP~$m$ is computed as $\Delta_m \mathbb{E} \{\| \mathbf{x}_m \|^2 \} = \Delta_m \sum_{k=1}^K \rho_{mk}$, where the scaling factor $\Delta_m \geq 1$ determines the inefficiency of the power amplifiers. In the second term in \eqref{eq:Ptotal}, $P_{m}$, models the power consumption of the transceiver chain connected to active APs and the traffic-independent power of the fronthaul connections and baseband processing. In the last part of \eqref{eq:Ptotal}, $P_{\mathrm{bt},m}$ (measured in Watt per bit/s) is the traffic-varying power consumption (of the fronthaul and baseband processing) that is proportional to the SE and system bandwidth $B$~Hz. When we activate an AP to improve the service, the power dissipation in the transceiver hardware in \eqref{eq:Ptotal} will increase, but the total transmit power might decrease thanks to the coherent combination of signals from multiple APs.
If the latter effect does not outweigh the former effect, it is better from an energy-efficiency perspective to keep the AP turned off, at least if it is still possible to satisfy the SE requirements of all users.

The total power consumption minimization problem that we want to solve is
\begin{equation} \label{Prob:TotalPowerOpt}
\begin{aligned}
& \underset{\{ \rho_{mk} \geq 0 \}, \mathcal{A} }{\mathrm{minimize}}
&& P_{\mathrm{total}} (\{ \rho_{mk} \}, \mathcal{A})  \\
& \,\textrm{subject to}
&&  R_{k} (\{ \rho_{mk} \}, \mathcal{A}) \geq \xi_{k}, \forall k,  \\
&&& \sum_{k=1}^K  \rho_{mk}  \leq P_{\max,m}, \forall m \in \mathcal{A},
\end{aligned}
\end{equation}
where $P_{\max,m}$ is the maximum downlink power of AP~$m$. The SE requirement of user~$k$ is denoted as $\xi_k$ [b/s/Hz] and thus the SE in \eqref{eq:DLRate} must be larger or equal to this number. Note that all the transmit power variables affect all the SEs due to mutual interference.

\begin{remark}
Similar optimization problems have been considered in \cite{vu2015energy, nguyen2018energy}, but under less practical conditions. The previous optimization problems are formulated for deterministic (or slowly fading) narrowband channels with perfect CSI , which require the transmitters, receivers, and propagation environment to be entirely static throughout the transmission of (infinitely) long codewords. 
The methods developed with such models are highly nontrivial to extend to practical fading wideband channels, where there will be channel estimation errors and the decisions of which APs to turn off must be done jointly over many narrowband subcarriers.
In contrast, we formulate problem~\eqref{Prob:TotalPowerOpt} based on the ergodic SE of fast fading channels, which is relevant in practical networks where the channels are rapidly changing and there is channel coding over multiple coherence intervals (spanning both over time and frequency). Since the optimization problems are formulated as a function of the large-scale fading coefficients, we find a solution that is appropriate for the entire wideband channel and for infinitely many small-scale fading realizations.
We stress that we are utilizing the specific features of Cell-free Massive MIMO to  compute SE expressions that take channel hardening, channel estimation errors, pilot contamination, and low-complexity linear precoding into account.
\end{remark}

 In many scenarios, the network only needs to activate a subset of the $M$ APs to deliver the required SE to the $K$ users, meaning that $\mathcal{A} \subseteq \{1, \ldots, M\}$. In order to study how many elements in $\mathcal{A}$ are needed, we set $\nu_k = 2^{\xi_k \tau_c /(\tau_c - \tau_p)} -1, \forall k$ and rewrite problem~\eqref{Prob:TotalPowerOpt} with SINR constraints as
\begin{equation} \label{Prob:TotalPowerSINR}
\begin{aligned}
& \underset{\{ \rho_{mk} \geq 0 \}, \mathcal{A} }{\mathrm{minimize}}
&& \sum_{m \in \mathcal{A}} \sum_{k=1}^K \Delta_m \rho_{mk} + \sum_{m \in \mathcal{A}} P_{\mathrm{hw},m}  \\
& \mathrm{subject\,\, to}
&&  \mathrm{SINR}_{k} (\{ \rho_{mk} \}, \mathcal{A})  \geq \nu_{k}, \forall k, \\
&&& \sum_{k=1}^K \rho_{mk} \leq P_{\max,m}, \forall m \in \mathcal{A},
\end{aligned}
\end{equation}
where the total hardware power consumption at AP~$m$, $P_{\mathrm{hw},m}$, is simplified from \eqref{eq:Ptotal} based on the fact that all the SINR constraints will be achieved with equality at an optimal solution \cite{shi2016sinr}:
\begin{equation} \label{eq:Phwm}
 P_{\mathrm{hw},m} = P_{m} + B \sum_{k=1}^K P_{\mathrm{bt},m} \xi_k.
\end{equation}
We have reduced the computational complexity of problem~\eqref{Prob:TotalPowerSINR} compared to \eqref{Prob:TotalPowerOpt} since the hardware power consumption \eqref{eq:Phwm} is now a constant, which transforms the objective function of problem~\eqref{Prob:TotalPowerSINR} from a nonlinear to a linear function. To further simplify the problem, we introduce the notations
\begin{align}
\mathbf{r}_{\mathcal{A}} = & \Bigg[ \sqrt{\Delta_{m_{1'}} \rho_{{m_{1'}}1}}, \ldots,  \sqrt{\Delta_{m_{|\mathcal{A}|}} \rho_{m_{|\mathcal{A}|}K}}, \sqrt{\sum_{m \in \mathcal{A}} P_{\mathrm{hw},m}} \Bigg]^T \notag \\
&\in \mathbb{C}^{|\mathcal{A}|K + 1} , \label{eq:r}\\
\mathbf{z}_{k\mathcal{A}} =& \left[\sqrt{z_{m_{1'}k}}, \ldots,\sqrt{z_{m_{|\mathcal{A}|}k}} \right]^T \in \mathbb{C}^{ |\mathcal{A}|}, \label{eq:zk} \\
%\end{align}
%\begin{align}
\mathbf{g}_{k\mathcal{A}} =& \left[ \sqrt{g_{1k}}, \ldots,  \sqrt{g_{m_{\mathcal{A}}k}} \right]^T \in \mathbb{C}^{ |\mathcal{A}|},\label{eq:gk} \\
\mathbf{U}_{\mathcal{A}} =& [\mathbf{u}_1, \ldots, \mathbf{u}_K]^T \in \mathbb{C}^{|\mathcal{A}| \times K}, \label{eq:U}\\
\mathbf{s}_{k\mathcal{A}} =& \Big[\sqrt{\nu_k}\big( \mathbf{g}_{k\mathcal{A}}^T \mathbf{u}_{t_1'}, \ldots, \mathbf{g}_{k\mathcal{A}}^T \mathbf{u}_{t_{|\mathcal{P}_k \setminus \{k\}|}'}, \| \mathbf{z}_{k\mathcal{A}} \circ \mathbf{u}_1 \|, \ldots, \notag \\
& \| \mathbf{z}_{k\mathcal{A}} \circ \mathbf{u}_K \|, \sigma_{\mathrm{DL}} \big) \Big]^T \in \mathbb{C}^{K + |\mathcal{P}_k|}, \label{eq:sk}
\end{align}
 where  $m_{1'},\ldots, m_{|\mathcal{A}|} $ are the members of the active AP set $\mathcal{A}$ (i.e., the indices of the active APs). In \eqref{eq:gk}, $g_{mk}$ is defined as $g_{mk}= N \gamma_{mk}$ for MRT precoding and $g_{mk}= (N - \tau_p) \gamma_{mk}$ for F-ZF precoding. The matrix $\mathbf{U}_{\mathcal{A}}$ in \eqref{eq:U} has the $k$-th column $\mathbf{u}_k = [\sqrt{\rho_{1k}}, \ldots, \sqrt{\rho_{m_{\mathcal{A}k}}}]^T$ and the $m$-th row is denoted as $\mathbf{u}_m'$. In \eqref{eq:sk}, $t_1', \ldots, t_{|\mathcal{P}_k \setminus \{k\}|}'$ are the indices of the users belonging to the set $\mathcal{P}_k \setminus \{k\}$, and $|\mathcal{P}_k|$ is the cardinality of the set $\mathcal{P}_k$. The operator $\circ$ denotes the Hadamard product. We can now obtain an equivalent epigraph representation of problem~\eqref{Prob:TotalPowerSINR} as
\begin{subequations} \label{Prob:NonConvexSOCP} 
\begin{alignat}{2}
& \underset{ \{ \rho_{mk} \geq 0 \}, \mathcal{A}, s_{\mathcal{A}} }{\mathrm{minimize}} \,\,\,
& &  s_{\mathcal{A}}\\
& \, {\mathrm{subject \,\, to}}
&&  \| \mathbf{r}_{\mathcal{A}} \| \leq  s_{\mathcal{A}}, \label{NonConvexSOCP:b}  \\
& & & \| \mathbf{s}_{k\mathcal{A}} \| \leq \mathbf{g}_{k\mathcal{A}}^T \mathbf{u}_{k\mathcal{A}}, \; \forall k = 1, \ldots, K, \label{NonConvexSOCP:c} \\
& & & || \mathbf{u}_{m}'|| \leq \sqrt{P_{\mathrm{max},m}}, \; \forall m \in \mathcal{A}. \label{NonConvexSOCP:d}
\end{alignat}
\end{subequations}
The auxiliary variable $s_{\mathcal{A}}$ moves the objective function  of problem~\eqref{Prob:TotalPowerSINR} to the first constraint in \eqref{NonConvexSOCP:b}. We observe that for a given $\mathcal{A}$, problem~\eqref{Prob:TotalPowerSINR} reduces to an SOC program, as previously shown in \cite{Ngo2017a, Nayebi2017a}. Hence, although \eqref{Prob:NonConvexSOCP} is non-convex, it can be solved by making an exhaustive search over all possible selections of $\mathcal{A}$ and solving each subproblem using convex optimization. Since at least one AP needs to be active if there is $K \geq 1$ users with non-zero SE requirements, there are $2^M - 1$ different selections of the APs that need to be considered in an exhaustive search. This naive approach to solving \eqref{Prob:TotalPowerSINR} will be very computationally costly even in a relatively small network.

\subsection{Globally Optimal Solution to the Total Power Minimization Problem}
Instead of making an exhaustive search, a global optimum to \eqref{Prob:NonConvexSOCP} can be achieved in a structured way by utilizing, for example, using the branch-and-bound approach \cite{bonami2008algorithmic}. That would result in a more efficient implementation but the computational complexity will still grow exponentially with the number of APs. However, it enables offline benchmarking in problems with up to tens of APs and users, as will be demonstrated numerically in Section~\ref{Sec:NumRes}.

Let the binary optimization variable $\alpha_m \in \{0,1 \}$ mathematically characterize the on/off activity of AP $m$. Instead of explicitly forcing the AP's transmit powers $\{\rho_{m1}, \ldots,  \rho_{mK}\}$ to zero when $\alpha_m = 0$, we can do it implicitly by 
replacing its maximum transmit power by $\alpha_m^2 P_{\max,m}$. This gives the original value $P_{\max,m}$ when the AP is active and is zero when the AP is turned off. This feature is exploited to formulate a mixed-integer SOC program as in Lemma~\ref{Theorem:MixedInSOCP}.
\begin{lemma} \label{Theorem:MixedInSOCP}
Consider the mixed-integer SOC program
\begin{subequations} \label{Prob:BnB} 
	\begin{alignat}{2}
	& \underset{ \{ \rho_{mk} \geq 0 \}, \{ \alpha_m \}, s }{\mathrm{minimize}} \,\,\,
	& &  s \\
	& \,\,\,{\mathrm{subject \,\, to}}
	&&  \| \mathbf{r} \| \leq  s, \label{BnB:b}  \\
	& & & \| \mathbf{s}_{k} \| \leq \mathbf{g}_{k}^T \mathbf{u}_{k}, \; \forall k = 1, \ldots, K, \label{BnB:c} \\
	& & & \| \tilde{\mathbf{u}}_{m}'\| \leq \alpha_m \sqrt{P_{\mathrm{max},m}}, \; \forall m = 1, \ldots, M, \label{BnB:d} \\
	&&& \alpha_m \in \{0,1 \}, \forall m = 1, \ldots, M,
	\end{alignat}
\end{subequations}
where $\tilde{\mathbf{u}}_{m}'$ is the $m$-th row of matrix $\widetilde{\mathbf{U}} = [\tilde{\mathbf{u}}_1, \ldots, \tilde{\mathbf{u}}_K] \in \mathbb{C}^{M \times K}$ and $\tilde{\mathbf{u}}_k = [\sqrt{\rho_{1k}}, \ldots, \sqrt{\rho_{Mk}}]^T \in \mathbb{C}^M, k = 1\ldots, K$. Moreover, the vectors $\mathbf{r}$ and $\mathbf{s}_k$ are defined as
\begin{align}
 \mathbf{r} =&  \Big[ \sqrt{\Delta_{1} \rho_{11}}, \ldots,  \sqrt{\Delta_{M} \rho_{MK}}, \alpha_1 \sqrt{P_{\mathrm{hw},1}}, \ldots,  \alpha_M \sqrt{P_{\mathrm{hw},M}} \Big]^T \notag \\
&\in \mathbb{C}^{MK + M} , \label{eq:rbnb}\\
%\end{align}
%\begin{align}
\mathbf{s}_k =& \Big[\sqrt{\nu_k}\big( \mathbf{g}_{k}^T \mathbf{u}_{t_1'}, \ldots, \mathbf{g}_{k}^T \mathbf{u}_{t_{|\mathcal{P}_k \setminus \{k\}|}'}, \| \mathbf{z}_{k} \circ \mathbf{u}_1 \|, \ldots,  \| \mathbf{z}_{k} \circ \mathbf{u}_K \|, \notag \\
& \sigma_{\mathrm{DL}} \big) \Big]^T  \in \mathbb{C}^{K + |\mathcal{P}_k|},\\
%\end{align}
%\begin{align}
 \mathbf{z}_{k} =& \left[\sqrt{z_{1k}}, \ldots,\sqrt{z_{Mk}} \right]^T \in \mathbb{C}^{ M},\\
 \mathbf{g}_{k} =& \left[ \sqrt{g_{1k}}, \ldots,  \sqrt{g_{Mk}} \right]^T \in \mathbb{C}^{M}.
\end{align}
Problems~\eqref{Prob:NonConvexSOCP} and \eqref{Prob:BnB} are equivalent in the sense that they have the same optimal transmit powers. If we denote by $\{\alpha_{m}^\ast \}$ an optimal solution to the binary variables $\{ \alpha_m \}$, which is obtained by solving problem~\eqref{Prob:BnB}, the optimal set of active APs in problem~\eqref{Prob:NonConvexSOCP} is
\begin{equation} \label{eq:ActiveAPv1}
\mathcal{A} = \left\{ m : \alpha_m^\ast = 1,  m \in \{1, \ldots, M \} \right\}.
\end{equation}
\end{lemma}
\begin{IEEEproof}
The binary variable $\alpha_m$ behaves as an indicator function which uniquely determines the activity of AP~$m$. When $\alpha_m = 0$, the related constraint~\eqref{BnB:d} is $\sum_{k=1}^K \rho_{mk} =  0$.  Since $\rho_{mk} \geq 0$, we obtain $\rho_{mk} = 0, \forall k=1, \ldots, K$. Alternatively, AP~$m$ will be turned off and it does not have any contribution to the total power consumption of the network as well as all terms that would have contained $\rho_{mk}$ in the SINR expression are missing in \eqref{BnB:c}. By contrast, when $\alpha_m = 1$, the related constraint~\eqref{BnB:d} becomes $\| \tilde{\mathbf{u}}_{m}'\| \leq \sqrt{P_{\mathrm{max},m}}$, 
which is a total transmit power constraint when AP~$m$ is in active mode as shown in~\eqref{Prob:NonConvexSOCP}. For that reason, finding $\{\alpha_{m}^{\ast} \}$ results is the same as optimizing the active APs set $\mathcal{A}$ in problem~\eqref{Prob:NonConvexSOCP} by utilizing \eqref{eq:ActiveAPv1}.
\end{IEEEproof}
The new binary variables provide the explicit link between the hardware and transmit power consumption, which is an important factor to obtain the global optimum to problem~\eqref{Prob:BnB}. A key reason that we can preserve the SOC structure, despite adding the new binary variables, is that the binary variables are not involved in the SINR constraints \eqref{BnB:c}. Instead there is an implicit connection via the zero maximum transmit power for inactive APs. This is different from the previous approaches, e.g., \cite{feng2017boost}, which also defined the on/off activity using $\alpha_m$ but then included it in the SINR expressions, leading to non-convex SINR constraints.

Problem~\eqref{Prob:BnB} is a mixed-integer SOC program on standard form, thus a globally optimal solution can be obtained using standard algorithms, for example, by using CVX \cite{cvx2015} in conjunction with the MOSEK solver \cite{Mosek}. This software applies the branch-and-bound approach \cite{bonami2008algorithmic} to deal with the binary variables. It is implemented in an iterative manner where the main cost of each iteration consisting three steps: finding a box, which gives a lower bound on the total power consumption, and splitting that box into the two new boxes; computing upper and lower bounds for the new generated boxes; and pruning boxes which cannot contain the optimum solution. The second step dominates the computational complexity of each iteration, while the third step decides the required number of iterations to reach the optimal solution. The following lemma provides an estimate of the computational complexity when solving problem~\eqref{Prob:BnB}.
\begin{lemma} \label{Lemma:ComplexityBnB}
By utilizing the standard interior-point method to solve a series of SOC programs, the computational complexity of the branch-and-bound approach to obtain a global optimum to problem~\eqref{Prob:NonConvexSOCP} is in the order of 
\begin{equation}
 \ln (\varepsilon^{-1}) \sum_{n=1}^{N_1} \sum_{ i \in \{ 0,1 \}} \mathcal{O} \left( C_i^{(n),\mathrm{ub}}\right) +  \mathcal{O} \left( C_i^{(n),\mathrm{lb}} \right) ,
\end{equation}
where $\varepsilon>0$ is the accuracy of solving SOC programs along the iterations.\footnote{For a given $\varepsilon$, the set of optimized variables is called $\varepsilon$-solution to an optimization problem if the objective function at this point is at most $\varepsilon$ away from the global optimum.} $N_1$  $(N_1 \leq 2^M -1)$ denotes the number of iterations needed for the branch-and-bound approach to reach an optimal solution. Moreover $C_i^{(n),\mathrm{ub}}$ and $C_i^{(n),\mathrm{lb}}$ denote the cost of computing the lower and upper bounds (see Appendix~\ref{Appendix:ComplexityBnB} for the definitions of these bounds), which are given by:
%\begin{figure*}
\begin{align}
 C_i^{(n),\mathrm{lb}}  =&  \sqrt{L_{i2}^{(n)}  + K^2  + L_{i1}^{(n)}K + Z^{(n)}} \times \notag \\
 & \left( \left(L_{i2}^{(n)} \right)^3 +      L_{i2}^{(n)}\sum_{k=1}^K \left| \mathcal{P}_k \right| + L_{i2}^{(n)} L_{i1}^{(n)} K^2 + Z^{(n)} K \right), \label{C0nub}\\
C_i^{(n), \mathrm{ub}} =& \big( U_i^{(n)} \big)^3 K^3  \sqrt{U_i^{(n)}K+ K^2 } \label{C1nlb}.
\end{align}
%\vspace*{-0.4cm}
%\hrulefill
%\end{figure*}
Here, $M_0^{(n-1)}$ and $M_1^{(n-1)}$ denote the number of APs already in active and sleep modes, respectively, which are obtained from the previous iteration. The initial values are $M_0^{(1)} = M_1^{(1)} = 0$. Moreover, $T^{(n)} = M - M_0^{(n-1)}$, $Q^{(n)} = M - M_0^{(n-1)} - M_1^{(n-1)}$, $Z^{(n)} = K\big( Q^{(n)} -1 \big)$ and the other parameters depend on the binary indices as
\begin{center}
	\begin{tabular}{| c|  c |c |}
		\hline
		Parameter & $i=0$ & $i=1$ \\ 
		\hline
		$L_{i1}^{(n)}$ & $M_1^{(n-1)}$ &  $M_1^{(n-1)} +1$ \\  
		\hline
		$L_{i2}^{(n)}$ & $\big(T^{(n)} -1 \big)K + Q^{(n)} $ & $T^{(n)}K + Q^{(n)}  $ \\  
		\hline  
		$U_{i}^{(n)}$ & $T^{(n)} -1$ & $T^{(n)}$ \\
		\hline
	\end{tabular}
\end{center}
\end{lemma}
\begin{IEEEproof}
The proof computes the computational complexity for solving SOC programs to achieve the upper and lower bounds that the branch-and-bound approach spends along iterations. A detailed derivation is provided in Appendix~\ref{Appendix:ComplexityBnB}.
\end{IEEEproof}
Lemma~\ref{Lemma:ComplexityBnB} shows that the computational complexity is the total cost of computing upper and lower bounds until reaching the global optimum. Even though the computational complexity per iteration varies as the change in both the optimization variables and the procedure needed in each iteration, \eqref{C0nub} and \eqref{C1nlb} can exhibit such features by using the big-$\mathcal{O}$ notation. In the worst case, the branch-and-bound approach has the same computational complexity as an exhaustive search over all $2^M-1$ boxes with possible subsets of active APs. With a proper bounding rule, the average computational complexity can be significantly reduced by pruning many boxes. Nevertheless, an exponential growth with $M$ is expected. In Section~\ref{Sec:NumRes}, we show that the branch-and-bound approach can find a globally optimal solution to a moderate-size network with $20$ APs.

\section{Two Suboptimal Algorithms With Lower Complexity} \label{Sec:SubSols}

Motivated by the high computational complexity of solving the total power minimization problem using Lemma~\ref{Theorem:MixedInSOCP}, we will now propose two algorithms that find good suboptimal solutions to problem~\eqref{Prob:NonConvexSOCP} with a tolerable computational complexity and enabling implementation in large Cell-free Massive MIMO networks.

\subsection{Utilizing Sparsity to Turn Off APs} \label{SubSec:Sparsity}
If the network does not need to turn on all the $M$ APs to provide the requested services from all the $K$ users, we know that many of the power variables will be zero. Hence, we can try to find the optimum AP subset by expressing \eqref{Prob:TotalPowerSINR} as a sparse reconstruction problem where we try to push many of the transmit power variables to become zero.
To this end, we first reformulate problem \eqref{Prob:TotalPowerSINR} as a mixed $\ell_2/\ell_0$-norm optimization problem.

\begin{lemma} \label{lemma:mixedL2L0}
	The original problem~\eqref{Prob:TotalPowerSINR} has the same optimal transmit powers as the following problem
	\begin{equation} \label{Prob:TotalPowerOptLp}
	\begin{aligned}
	& \underset{\{ \rho_{mk} \geq 0 \} }{\mathrm{minimize}}
	&&   \sum_{ m=1}^M \Delta_m \| \pmb{\rho}_{m} \|^{2}  + \mathbbm{1}_m (\pmb{\rho}_{m}) P_{\mathrm{hw},m}\\
	& \,\,\textrm{subject to}
	& &  \| \mathbf{s}_{k\mathcal{A}_M} \| \leq \mathbf{g}_{k\mathcal{A}_M}^T \mathbf{u}_{k\mathcal{A}_M}, \; \forall k = 1,\ldots, K,\\
	& & & || \mathbf{u}_{m}'|| \leq \mathbbm{1}_m (\pmb{\rho}_{m}) \sqrt{P_{\mathrm{max},m}}, \; \forall m = 1, \ldots, M,
	\end{aligned}
	\end{equation}
	where $\pmb{\rho}_m = [\sqrt{\rho_{m1}}, \ldots, \sqrt{\rho_{mK}}]^T \in \mathbb{C}^K$ and each function $\mathbbm{1}_m (\pmb{\rho}_{m})$  is defined based on the transmit powers of AP~$m$ as
	\begin{equation} \label{eq:IndxFunc}
	\mathbbm{1}_m (\pmb{\rho}_{m}) = \begin{cases}
	1, & \mbox{if } \| \pmb{\rho}_{m} \| > 0, \\
	0, & \mbox{if } \| \pmb{\rho}_{m} \| = 0.
	\end{cases}
	\end{equation}
Moreover, if we denote by $\{ \rho_{mk}^{\ast} \}$ the optimal set of all transmit powers to \eqref{Prob:TotalPowerOptLp}, then the set
\begin{equation} \label{eq:AsetV1}
\mathcal{A} = \left\{ m : \| \pmb{\rho}_m^{\ast} \| > 0, m \in \{1, \ldots, M \} \right\}
\end{equation}
is the optimal set of active APs to problem~\eqref{Prob:TotalPowerSINR}.
\end{lemma}
\begin{IEEEproof}
When AP~$m$ is in sleep mode, it assigns zero transmit power to all users (i.e., $\rho_{mk} = 0, \forall k = 1, \ldots, K$). This AP has no contribution to the objective function of problem~\eqref{Prob:TotalPowerOptLp} due to $ \| \pmb{\rho}_m \| = 0$ and thus we can make the definition of $\mathbbm{1}_m (\pmb{\rho}_{m})$ as in \eqref{eq:IndxFunc}. The optimal set of active APs is defined based on the group-sparsity structure as in \eqref{eq:AsetV1}. 
\end{IEEEproof}
Lemma~\ref{lemma:mixedL2L0} shows that we do not need to define separate variables for optimizing the active APs set~$\mathcal{A}$, but we can implicitly determine if AP $m$ is active or not by checking if $\| \pmb{\rho}_m \| > 0 $ or $\| \pmb{\rho}_m \| = 0 $. The reformulated problem~\eqref{Prob:TotalPowerOptLp} reduces the number of optimization variables compared with \eqref{Prob:BnB}, and in particular, all the optimization variables are now continuous. Nevertheless, problem~\eqref{Prob:TotalPowerOptLp} is still non-convex due to $\ell_0$-norm in the second part of the objective function. However, we can relax $\ell_0$-norm to an $\ell_p$-norm for some $0<p<1$. This is a standard relaxation technique that retains sparsity and we stress that it also gives better sparsity than an $\ell_1$-norm relaxation (cf.~Figs.~$2$ and $3$ in \cite{zhang2015survey} for illustrations).\footnote{Strictly speaking, a value $p\in [0,1)$ does not lead to norm since the subadditive property is not satisfied \cite{Horn2013a}, but the ``norm" terminology has anyway been used for many years and we adopt this convention.} Therefore, we adopt the $\ell_p$-norm optimization to obtain a relaxation of problem~\eqref{Prob:TotalPowerOptLp} as\footnote{From the range of the considered $\ell_p$-norms, the condition $0 < \tilde{p}/2 < 1$ as in \eqref{Prob:TotalPowerOptLpv0} leads to $0 < \tilde{p} < 2$.} 
\begin{equation} \label{Prob:TotalPowerOptLpv0}
\begin{aligned}
&\,\, \underset{\{ \rho_{mk} \geq 0 \} }{\mathrm{minimize}}
&&   \left( \sum_{ m=1}^M \left(\Delta_m^{2/\tilde{p}} \| \pmb{\rho}_{m} \|^{2} \right)^{\tilde{p}/2}  + P_{\mathrm{hw},m}^{\tilde{p}/2} \right)^{2/\tilde{p}}\\
& \,\,\mathrm{subject \,\, to}
& &  \| \mathbf{s}_{k \mathcal{A}_M} \| \leq \mathbf{g}_{k\mathcal{A}_M}^T \mathbf{u}_{k\mathcal{A}_M}, \; \forall k = 1,\ldots, K,\\
& & & || \mathbf{u}_{m}'|| \leq \sqrt{P_{\mathrm{max},m}}, \; \forall m = 1, \ldots, M.
\end{aligned}
\end{equation}
The objective function of problem~\eqref{Prob:TotalPowerOptLpv0} treats every vector $\pmb{\rho}_{m}$ as an entity in $\Delta_m^{2/\tilde{p}} \| \pmb{\rho}_{m} \|^{2}$ when seeking a sparse solution. We will utilize this group-sparse property of the transmit power coefficients $\pmb{\rho}_m$ to solve problem~\eqref{Prob:TotalPowerOptLpv0} in a novel way, which differs from previous works that considered element-based  \cite{candes2008enhancing} or beamforming-vector-based sparsity \cite{luo2014downlink}. Even though  problem~\eqref{Prob:TotalPowerOptLpv0} remains non-convex after the norm relaxation, we can find a stationary point by adapting the iteratively reweighted least squares approach from \cite{ba2014}, that was originally developed for problems with component-wise sparsity. Specifically, after removing the exponent $2/\tilde{p}$ and the hardware power consumption in the objective function, problem~\eqref{Prob:TotalPowerOptLpv0} can be recast as
\begin{equation} \label{Prob:TotalPowerOptLpv1}
\begin{aligned}
& \underset{\{ \rho_{mk} \geq 0 \} }{\mathrm{minimize}}
&&    \sum_{ m=1}^M \Delta_m \| \pmb{\rho}_{m} \|^{\tilde{p}} \\
& \,\,\textrm{subject to}
& &  \| \mathbf{s}_{k \mathcal{A}_M} \| \leq \mathbf{g}_{k\mathcal{A}_M}^T \mathbf{u}_{k\mathcal{A}_M}, \; \forall k = 1,\ldots, K,\\
& & & || \mathbf{u}_{m}'|| \leq \sqrt{P_{\mathrm{max},m}}, \; \forall m = 1, \ldots, M.
\end{aligned}
\end{equation}
By noting that the group-sparse property implies the support of vector $\pmb{\rho}_m$ is the empty set, we can provide an iterative algorithm obtaining a stationary solution to problem~\eqref{Prob:TotalPowerOptLpv1}.
\begin{theorem} \label{Theorem:Sparsity}
	%Suppose problem~\eqref{Prob:TotalPowerOptLpv1} is feasible.
	Since the feasible set is convex, we can construct an iterative algorithm that starts with the given initial weight values $a_m^{(0)} = 1, \forall m=1, \ldots, M,$ and in iteration $n=1,2,\ldots$ solves the SOC program
	\begin{equation} \label{Prob:TotalPowerOptv3}
	\begin{aligned}
	& \underset{\{ \rho_{mk} \geq 0 \} }{\mathrm{minimize}}
	&&   \sum_{ m=1}^M a_m^{(n-1)} \| \pmb{\rho}_{m} \|^{2}\\
	& \,\,\textrm{subject to}
	& &  \| \mathbf{s}_{k \mathcal{A}_M} \| \leq \mathbf{g}_{k\mathcal{A}_M}^T \mathbf{u}_{k\mathcal{A}_M}, \; \forall k = 1,\ldots, K,\\
	& & & || \mathbf{u}_{m}'|| \leq \sqrt{P_{\mathrm{max},m}}, \; \forall m = 1, \ldots, M,
	\end{aligned}
	\end{equation}
	to yield the solution $\{ \pmb{\rho}_{m}^{\ast, (n)} \}$, for which
	\begin{equation}
	\pmb{\rho}_{m}^{\ast,(n)} = \left[\sqrt{\rho_{m1}^{\ast,(n)}}, \ldots, \sqrt{\rho_{mK}^{\ast,(n)}} \right]^T \in \mathbb{C}^K,
	\end{equation}
	is the optimal transmit powers for AP~$m$ at iteration~$n$. After that, the weight values are updated for the next iteration as
	\begin{equation} \label{eq:weightv1}
	a_m^{(n)} = \frac{\Delta_m \tilde{p}}{2}  \left( \| \pmb{\rho}_{m}^{\ast, (n)} \|^2 + \epsilon_n^2 \right)^{ \frac{\tilde{p}}{2} - 1},
	\end{equation}
where $\epsilon_n$ is a sufficiently small positive damping constant with $\epsilon_n \leq \epsilon_{n-1}$. When $\lim_{n \rightarrow \infty } \epsilon_n = 0$, the proposed iterative process exhibits the properties below:
	\begin{enumerate}
		\item The objective function of problem~\eqref{Prob:TotalPowerOptLpv1} reduces after each iteration until reaching a fixed point, which is a stationary point of problem~\eqref{Prob:TotalPowerOptLpv1}.
		\item If an arbitrary AP~$m$ has zero transmit power at the optimum of iteration $n$, this AP will have zero transmit power in all the following iterations.
	\end{enumerate} 
\end{theorem}
\begin{IEEEproof}
The proof is based on the convergence property of the iteratively weighted least squares approach that has been adapted to our framework. The detailed proof is available in Appendix~\ref{Appendix:Sparisity}.
\end{IEEEproof}

Theorem~\ref{Theorem:Sparsity} guarantees a monotonically decreasing objective function and the main computational cost is to solve \eqref{Prob:TotalPowerOptv3} in each iteration. The iterative process reaches a stationary point to problem~\eqref{Prob:TotalPowerOptLpv1}. The second property supports turning off APs along iterations. The damping constant $\epsilon_n >0$ is introduced to cope with a numerical issue that can appear when updating the weight values \eqref{eq:weightv1}, i.e., $a_m^{(n)} \rightarrow \infty$ when $\| \pmb{\rho}_{m}^{\ast, (n)} \| \rightarrow 0$. Even though the convergence properties in Theorem~\ref{Theorem:Sparsity} are proved by the descending of $\epsilon_n$ along iterations, a sufficiently small constant value also works well in the simulations as reported in \cite{chartrand2008}. The stopping criterion can be selected by comparing two consecutive iterations. For a given accuracy $\varepsilon >0$, we can verify if
%\begin{equation}
$\delta \leq \varepsilon$,
%\end{equation}
where $\delta$ is the difference of the objective function to problem~\eqref{Prob:TotalPowerOptv3}:
\begin{equation} \label{eq:delta}
\delta = \left| \sum_{m=1}^M \Delta_m \|\pmb{\rho}_{mk}^{\ast,(n-1)} \|^{\tilde{p}}  - \sum_{m=1}^M \Delta_m  \| \pmb{\rho}_{mk}^{\ast,(n)}\|^{\tilde{p}} \right|.
\end{equation}
The stationary point achieved by Theorem~\ref{Theorem:Sparsity} may not be a globally optimal solution to problem~\eqref{Prob:TotalPowerOptLp} due to the norm relaxation and the inherent non-convexity. Consequently, we will not use the solution from Theorem~\ref{Theorem:Sparsity} as the final solution but instead as an indication of which APs to further turn off. More precisely, we compute the transmit power that the APs utilize at the solution from Theorem~\ref{Theorem:Sparsity} and reorder the APs in increasing power order.\footnote{We have implemented other possible orderings, for example, based on the total transmit power per AP or the relative maximum
	received power allocated to the users. For brevity, we are only considering the one that gave the best results.}

Let us denote by $\{\rho_{mk}^{\ast} \}, \forall m=1, \ldots M, k=1, \ldots,K,$ the optimized transmit powers obtained by Theorem~\ref{Theorem:Sparsity}, for which a new parameter $\theta_m$ standing for the contribution of AP~$m$ is defined. Specifically,  $\theta_m$ is the total received power of the $K$ users that is transmitted by AP~$m$ as:
\begin{equation} \label{eq:Thetam}
\theta_{m} = \sum_{k =1}^K \rho_{mk}^{\ast} \left| \mathbb{E} \{\mathbf{h}_{mk}^H \mathbf{w}_{mk}\} \right|^2 = N \sum_{k =1}^K \rho_{mk}^{\ast} \beta_{mk}.
\end{equation}
In order to classify the contribution of each AP in $\mathcal{A}^{\ast}$ to provide the required SINRs, 
we define a heuristic ascending order as\footnote{Note that APs that were inactive at the solution obtained from Theorem~\ref{Theorem:Sparsity} are still considered. Since the numerical precision is limited, these APs will be assigned extremely small but non-zero power, which leads to a unique ordering in \eqref{eq:Order}.}
\begin{equation} \label{eq:Order} 
\theta_{1'} \leq \theta_{2'} \leq \ldots \leq \theta_{M'},
\end{equation}
where $\{1', \ldots, M'\}$ is a permutation of $\{1, \ldots, M\}$. 
We will now decide how many APs to utilize and keep only those with the largest $\theta$-values using the ordering in \eqref{eq:Order}. We further compute an auxiliary variable
\begin{equation} \label{eq:sast}
s^{\ast} = \sqrt{P_{\mathrm{total}}\left( \{\rho_{mk}^\ast \}, \mathcal{A}^{\ast} \right)}.
\end{equation}
We begin by defining a range $[M_{\mathrm{low}}, M_{\mathrm{up}}]$, with the condition $M_{\mathrm{up}} - M_{\mathrm{low}} \geq 1$. Specifically, the initial values are  $M_{\mathrm{low}} = 1$ and $M_{\mathrm{up}} = M$ , then we compute the middle point at iteration $\tilde{n}$ as
\begin{equation} \label{eq:mdoubleprime}
\tilde{m}^{(\tilde{n})} = \left \lfloor (M_{\mathrm{low}} + M_{\mathrm{up}})/2 \right \rfloor,
\end{equation}
where $\lfloor \cdot \rfloor$ denotes the floor function. We now reorder the AP indices according to \eqref{eq:Order} and consider setting the first $\tilde{m}^{(\tilde{n})}-1$ APs into sleep mode. Then, the active APs set is given by $\mathcal{A} = \mathcal{A}_{\tilde{n}} = \{\tilde{m}^{(\tilde{n})}, \ldots, M \},$ which has the cardinality $|\mathcal{A}_{\tilde{n}}| = M-\tilde{m}^{(\tilde{n})} +1$. We now solve the following SOC program:
\begin{equation} \label{Prob:ConvexSOCPv1} 
\begin{aligned}
& \underset{ \{ \rho_{mk} \geq 0 \}, s_{\mathcal{A}_{\tilde{m}^{(\tilde{n})}}} }{\mathrm{minimize}}
& &  s_{\mathcal{A}_{\tilde{m}^{(\tilde{n})}}}\\
& \,\,\,{\mathrm{subject \,\, to}}
&&  \| \mathbf{r}_{\mathcal{A}_{\tilde{m}^{(\tilde{n})}}} \| \leq  s_{\mathcal{A}_{\tilde{m}^{(\tilde{n})}}}, \\
& & & \| \mathbf{s}_{k\mathcal{A}_{\tilde{m}^{(\tilde{n})}}} \| \leq \mathbf{g}_{k\mathcal{A}_{\tilde{m}^{(\tilde{n})}}}^T \mathbf{u}_{k\mathcal{A}_{\tilde{m}^{(\tilde{n})}}}, \; \forall k = 1,\ldots, K,\\
& & & || \mathbf{u}_{m}'|| \leq \sqrt{P_{\mathrm{max},m}}, \; \forall m = \tilde{m}^{(\tilde{n})}, \ldots M',
\end{aligned}
\end{equation}
where $s_{\mathcal{A}_{\tilde{m}^{(\tilde{n})}}}$ is an upper bound defined by the sublevel set in the epigraph representation of problem~\eqref{Prob:TotalPowerSINR} when the $M - \tilde{m}^{(\tilde{n})} +1$ APs are in active mode. From the solution to problem~\eqref{Prob:ConvexSOCPv1}, the new upper or lower bounds on the number of inactive APs are updated as
\begin{equation} \label{eq:UpdateM}
\begin{cases}
M_{\mathrm{low}} = \tilde{m}^{(\tilde{n})}, & \mbox{if \eqref{Prob:ConvexSOCPv1} is feasible  and }   s_{\mathcal{A}_{\tilde{m}^{(\tilde{n})}}}^\ast <  s^\ast,\\
M_{\mathrm{up}} = \tilde{m}^{(\tilde{n})},& \mbox{otherwise},
\end{cases}
\end{equation}
where $s_{\mathcal{A}_{\tilde{m}^{(\tilde{n})}}}^\ast$ is the solution to \eqref{Prob:ConvexSOCPv1} at iteration~$\tilde{n}$. Notice from \eqref{eq:UpdateM} that when $M_{\mathrm{low}}$ is updated, the current optimal transmit powers and active APs set will be stored. Moreover, we update $s^\ast = s_{\mathcal{A}_{\tilde{m}^{(\tilde{n})}}}^\ast$. This iterative process will be executed until $M_{\mathrm{up}} - M_{\mathrm{low}} = 1$ as summarized in Algorithm~\ref{Algorithm:Sparsity}. If we assume problem~\eqref{Prob:NonConvexSOCP} has an optimal solution, then Algorithm~\ref{Algorithm:Sparsity} can always keep track of the best feasible point among those that are observed when running this algorithm, thanks to the condition $s_{\mathcal{A}_{\tilde{m}^{(\tilde{n})}}}^\ast <  s^\ast$. We further obtain the computational complexity of Algorithm~\ref{Algorithm:Sparsity} as in Lemma~\ref{lemma:ComplexitySparsity}.
\begin{algorithm}[t]
	\caption{Selecting how many APs to turn off with sparsity support} \label{Algorithm:Sparsity}
	\textbf{Input}: Large-scale fading coefficients $\beta_{mk}, \forall, m,k$; Maximum power levels $P_{l,}, \forall l,k$; Initial weight values $a_m^{(0)} = 1, \forall m$; Set iteration index $n=0$ and $\delta = \infty$; Set $M_{\mathrm{low}} = 1$ and $M_{\mathrm{up}} = M$.
	\begin{itemize}
		\item[1.] \textbf{while} $\delta > \epsilon$ \textbf{do}
		\begin{itemize}
			\item[1.1.] Set $n=n+1$.
			\item[1.2.] Solve problem~\eqref{Prob:TotalPowerOptv3} by using the previous weight values $a_m^{(n-1)}, \forall m,$ to obtain the optimal transmit powers $\rho_{mk}^{\ast,(n)}, \forall m,k.$
			\item[1.3.] Update the weight values $a_{m}^{(n)}$ by using \eqref{eq:weightv1}.
			\item[1.4.] Compute the stopping value $\delta$ in \eqref{eq:delta}.
		\end{itemize}
		\item[2.] \textbf{End while}
		\item[3.] Set $\rho_{mk}^\ast = \rho_{mk}^{\ast,(n)},\forall l,k,$ define $\mathcal{A}^\ast,$ and compute $s^{\ast}$ as in \eqref{eq:sast}; Compute $\theta_m , \forall m,$ as in \eqref{eq:Thetam}; Define the ascending order in \eqref{eq:Order}. Set $\tilde{n}=1$.
		\item[4.] \textbf{while} $M_{\mathrm{up}} - M_{\mathrm{low}} > 1$ \textbf{do}
		\begin{itemize}
			\item[4.2.] Compute $\tilde{m}^{(\tilde{n})}$ as in \eqref{eq:mdoubleprime} and then solve problem~\eqref{Prob:ConvexSOCPv1}.
			\item[4.3.] \textbf{If} problem~\eqref{Prob:ConvexSOCPv1} is feasible and $s_{\mathcal{A}_{\tilde{m}^{(\tilde{n})}}}^\ast <  s^\ast$, \textbf{then}: Set $M_{\mathrm{low}} = \tilde{m}^{(\tilde{n})}$; Update the current optimal solution $\rho_{mk}^{\ast}= \rho_{mk}^{\ast,(\tilde{n})},\forall m,k,\mathcal{A}^\ast = \mathcal{A}_{\tilde{m}^{(\tilde{n})}}^\ast,$ and $s^\ast = s_{\mathcal{A}_{\tilde{m}^{(\tilde{n})}}}^\ast$. \textbf{Otherwise}, Set  $M_{\mathrm{up}} = \tilde{m}^{(\tilde{n})}$.
			\item[4.4.] Set $\tilde{n}= \tilde{n}+1$.
		\end{itemize}
		\item[5.] \textbf{End while}
	\end{itemize}
	\textbf{Output}: The optimized transmit powers: $\rho_{mk}^{\mathrm{opt}} = \rho_{mk}^{\ast},\forall m,k,$ and active APs set $\mathcal{A}^\ast$. 
\end{algorithm}
\begin{lemma} \label{lemma:ComplexitySparsity}
	By using the standard interior-point method, the complexity order of Algorithm~\ref{Algorithm:Sparsity} to obtain the given $\varepsilon$-accuracy is
	\begin{multline} \label{eq:ComplexitySparse}
	\ln\big(\varepsilon^{-1} \big) \Big(  \mathcal{O} \left( N_2 M^3 K^3 \sqrt{MK + K^2} \right) +  \\
	\sum_{\tilde{n}_1=1}^{N_3}   \sum_{\tilde{n}_2=1}^{N_3} \mathcal{O}\Big(\sqrt{ \big(M- \tilde{m}^{(\tilde{n}_1)}\big)K + K^2 } \big( M- \tilde{m}^{(\tilde{n}_2)}\big)^3 K^3\Big) \Big),
	\end{multline}
	where $N_2$ is the number of iterations needed for finding the group sparsity support. $N_3$ is the number of iterations needed for the turnoff APs which satisfies $N_3 \leq \lceil \log_2 (M+1) \rceil$.
\end{lemma}
\begin{IEEEproof}
	The computational complexity order is obtained as in \eqref{eq:ComplexitySparse} by determining the costs of computing the two main iterative tasks: The optimal transmit powers when the $M$ APs are in active mode and a better local optimum by turning off APs. The main computational complexity in each task lies in solving SOC programs with the similar algebra in the proof of Lemma~\ref{Lemma:ComplexityBnB}, but we are now only considering the transmit powers as optimization variables.
	
	For the second task, an upper bound on the number of iterations needed for the turnoff APs stage is obtained by taking the lower and upper bound on the number of inactive APs, and dividing it in half, we will take the largest of those two intervals. The following inequality holds at iteration~$\tilde{n}$:
	\begin{equation} \label{eq:Bound}
	M_{\mathrm{up}} - M_{\mathrm{low}} \leq \frac{M-1}{2^{\tilde{n}}} + \frac{1}{2^{\tilde{n}-1}}.
	\end{equation}
	In the right hand-side of \eqref{eq:Bound}, the first part stands for the error bound of splitting intervals, while the other is for rounding the middle point of each interval. Algorithm~$1$ terminates when $M_{\mathrm{up}} - M_{\mathrm{low}} =1$, so \eqref{eq:Bound} becomes
	%\begin{equation}
	$2^{\tilde{n}} \leq M+1$
	%\end{equation}
	and therefore an upper bound of the number of iterations is obtained as in the lemma.
\end{IEEEproof} 

Algorithm~\ref{Algorithm:Sparsity} has a computational complexity in the order of
\begin{equation}
\mathcal{O}\left( N_2 M^{3.5} K^{3.5} + N_3^2 \left( M- \underset{  \tilde{m}^{(\tilde{n})} }{ \mathrm{argmax} } \, \, \tilde{m}^{(\tilde{n})}\right)^{3.5} K^{3.5} \right).
\end{equation}
In comparison to the branch-and-bound approach, the computational complexity per iteration reduces since only one SOC program is solved in each iteration. Moreover, for large-scale networks, the number of iterations $(N_2 + N_3)$ required in Algorithm~\ref{Algorithm:Sparsity} is expected to be much less than with the branch-and-bound approach. 

\subsection{Total Transmit Power Minimization and Turnoff APs} \label{SubSec:TotalTransmitPow}
The second low-complexity algorithm is obtained by optimizing the transmit powers only once for the case when all APs are turned on (i.e., $\mathcal{A} = \mathcal{A}_M = \{1, \ldots, M \}$) and then use this solution to decide in which order that APs should be turned off. Then problem~\eqref{Prob:NonConvexSOCP} becomes
\begin{equation} \label{Prob:ConvexSOCP} 
\begin{aligned}
& \underset{ \{ \rho_{mk} \geq 0 \}, s_{\mathcal{A}_M} }{\mathrm{minimize}}
& &  s_{\mathcal{A}_M}\\
& {\mathrm{subject \,\, to}}
&&  \| \mathbf{r_{\mathcal{A}_M}} \| \leq  s_{\mathcal{A}_M}, \\
& & & \| \mathbf{s}_{k \mathcal{A}_M} \| \leq \mathbf{g}_{k\mathcal{A}_M}^T \mathbf{u}_{k\mathcal{A}_M}, \; \forall k = 1,\ldots, K,\\
& & & || \mathbf{u}_{m}'|| \leq \sqrt{P_{\mathrm{max},m}}, \; \forall m = 1, \ldots M,
\end{aligned}
\end{equation}
which is an SOC program and we can obtain the optimal solution in polynomial time. We use the solution to this problem to order the APs according to \eqref{eq:Thetam} and \eqref{eq:Order}, and then follow the same procedure as in Algorithm~\ref{Algorithm:Sparsity} to determine how many APs should be active. This results in Algorithm~\ref{Algorithm:TotalTransPower} and the computational complexity order is obtained from Lemma~\ref{lemma:ComplexitySparsity} by setting $N_2 =1$.
	\begin{algorithm}[t]
	\caption{Total transmit power minimization and turnoff APs} \label{Algorithm:TotalTransPower}
	\textbf{Input}:  Large-scale fading coefficients $\beta_{mk}, \forall m,k$; Maximum power levels $P_{\max,m}, \forall m$; Set up $M_{\mathrm{low}} = 0$ and $M_{\mathrm{up}} = M$; Set $\mathcal{A} = \mathcal{A}_M = \{1,\ldots,M \}$.
	\begin{itemize}
		\item[1.] Solve problem~\eqref{Prob:ConvexSOCP} to obtain $\rho_{mk}^\ast, \forall m,k$.
		\item[2.] Compute $\theta_m, \forall m, =1, \ldots, M,$ and sort them in the ascending order as \eqref{eq:Order}; Set $\tilde{n}=1$.
		\item[3.] Perform the turnoff APs similar to Algorithm~\ref{Algorithm:Sparsity}.
	\end{itemize}
	\textbf{Output}: The optimized transmit powers: $\rho_{mk}^{\mathrm{opt}} = \rho_{mk}^{\ast},\forall m,k,$ and active APs set $\mathcal{A}^\ast$. 
\end{algorithm}
It means that Algorithm~\ref{Algorithm:TotalTransPower} has lower complexity than Algorithm~\ref{Algorithm:Sparsity} since it only solves one SOC program to determine the AP ordering. In more detail, the computational complexity of Algorithm~\ref{Algorithm:TotalTransPower} increases roughly as a polynomial of the optimization variables, i.e., $\mathcal{O}(M^{3.5} K^{3.5})$. 

\begin{figure*}[t]
	\begin{minipage}{0.48\textwidth}
		\centering
		\includegraphics[trim=0.5cm 0.0cm 1cm 0.5cm, clip=true, width=3.0in]{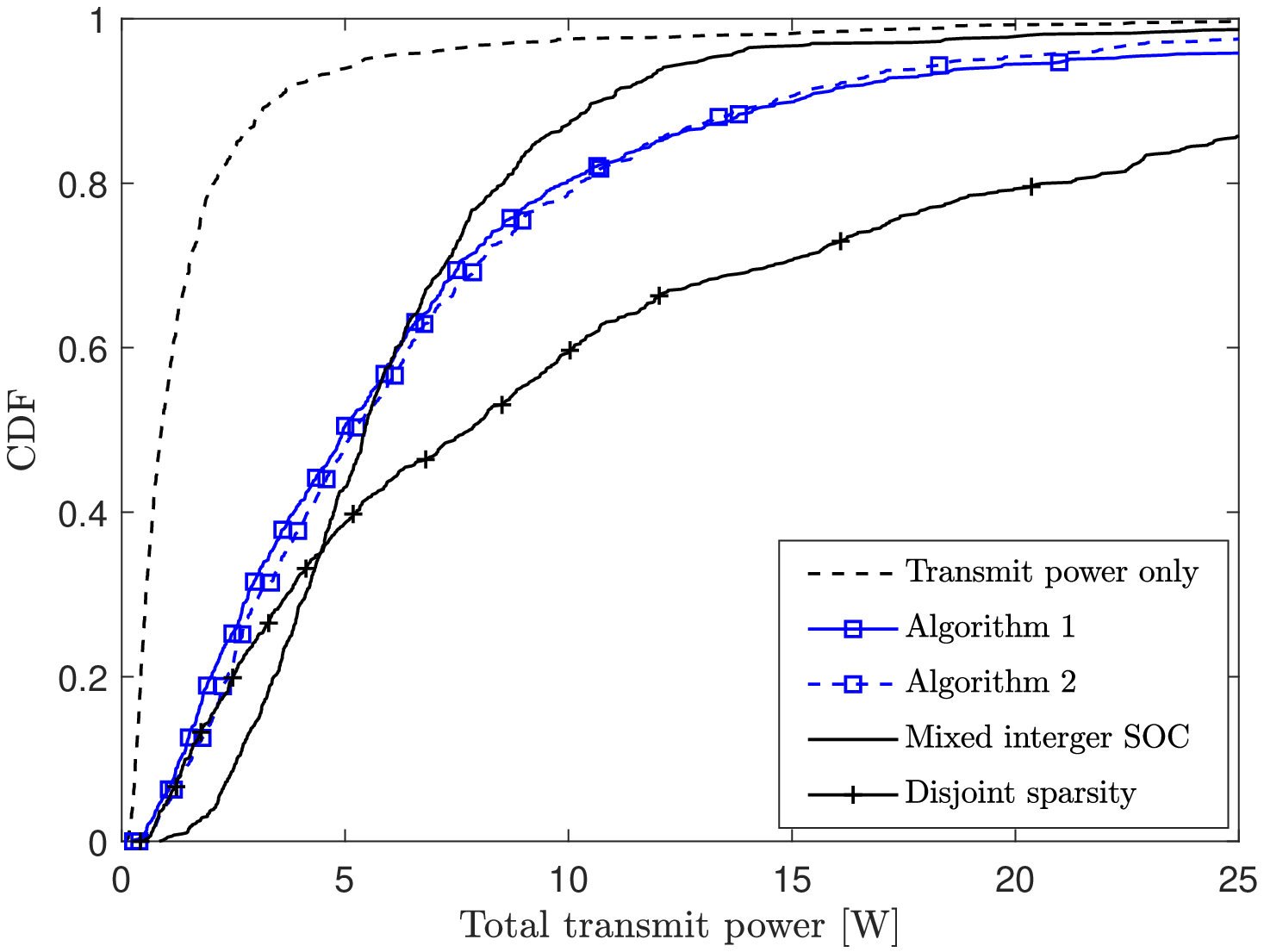} \vspace*{-0.25cm}
		\caption{The CDF of the total transmit power [W] with $M=20, K=20,$ and MRT precoding.}
		\label{FigTransmitPowerMRT}
		\vspace*{-0.4cm}
	\end{minipage}
	\hfill
	\begin{minipage}{0.48\textwidth}
		\centering
		\includegraphics[trim=0.5cm 0.0cm 1cm 0.5cm, clip=true, width=3.0in]{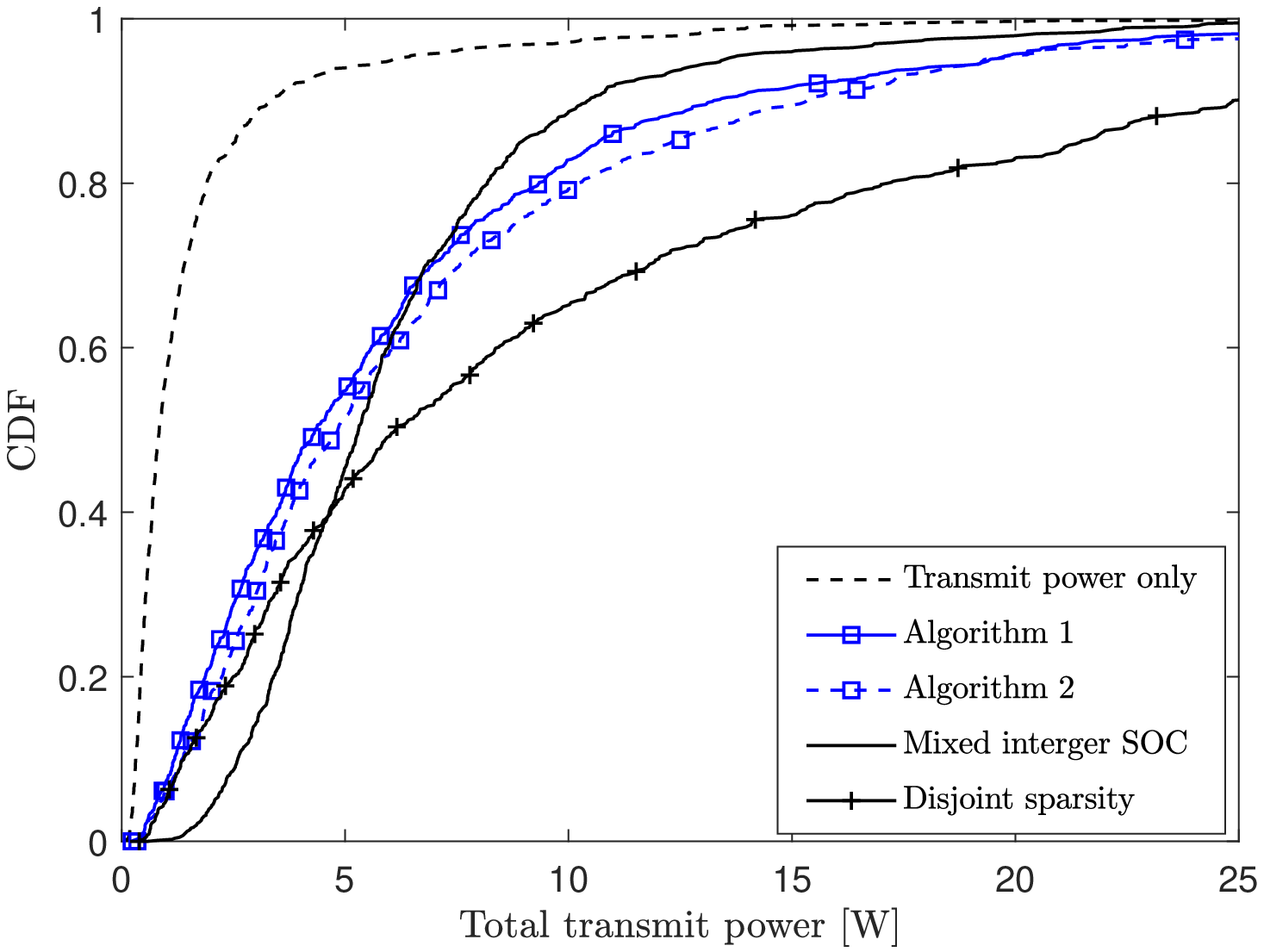} \vspace*{-0.25cm}
		\caption{The CDF of the total transmit power [W] with $M=20, K=20,$ and F-ZF precoding.}
		\label{FigTransmitPowerZF}
		\vspace*{-0.4cm}
	\end{minipage}
\end{figure*}
\begin{figure*}[t]
	\begin{minipage}{0.48\textwidth}
		\centering
		\includegraphics[trim=0.5cm 0.0cm 1cm 0.5cm, clip=true, width=3.0in]{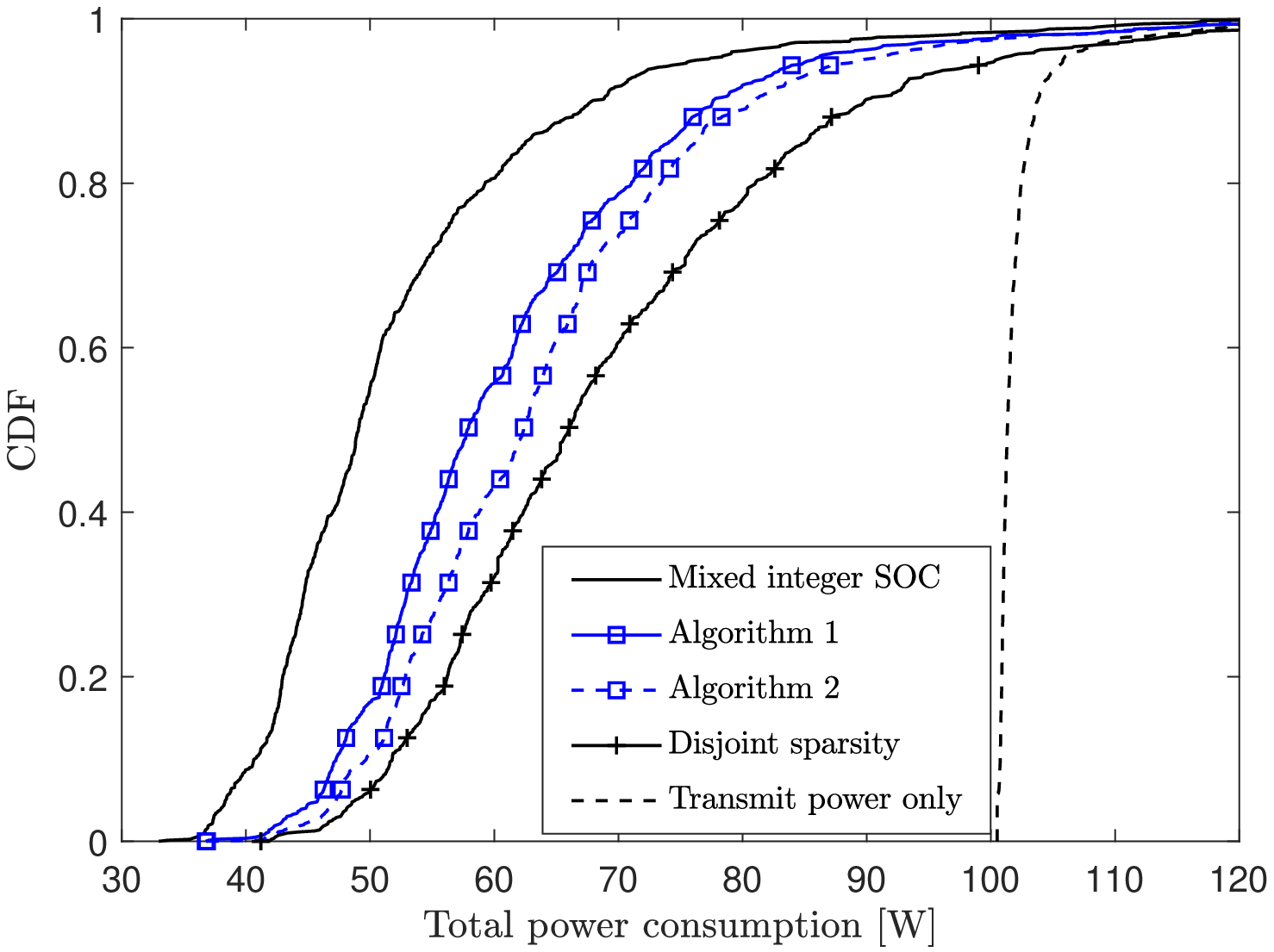} \vspace*{-0.25cm}
		\caption{The CDF of the total power consumption [W] with $M=20, K=20,$ and MRT precoding.}
		\label{FigTotalPowerMRT}
		\vspace*{-0.4cm}
	\end{minipage}
	\hfill
	\begin{minipage}{0.48\textwidth}
		\centering
		\includegraphics[trim=0.5cm 0.0cm 1cm 0.5cm, clip=true, width=3.0in]{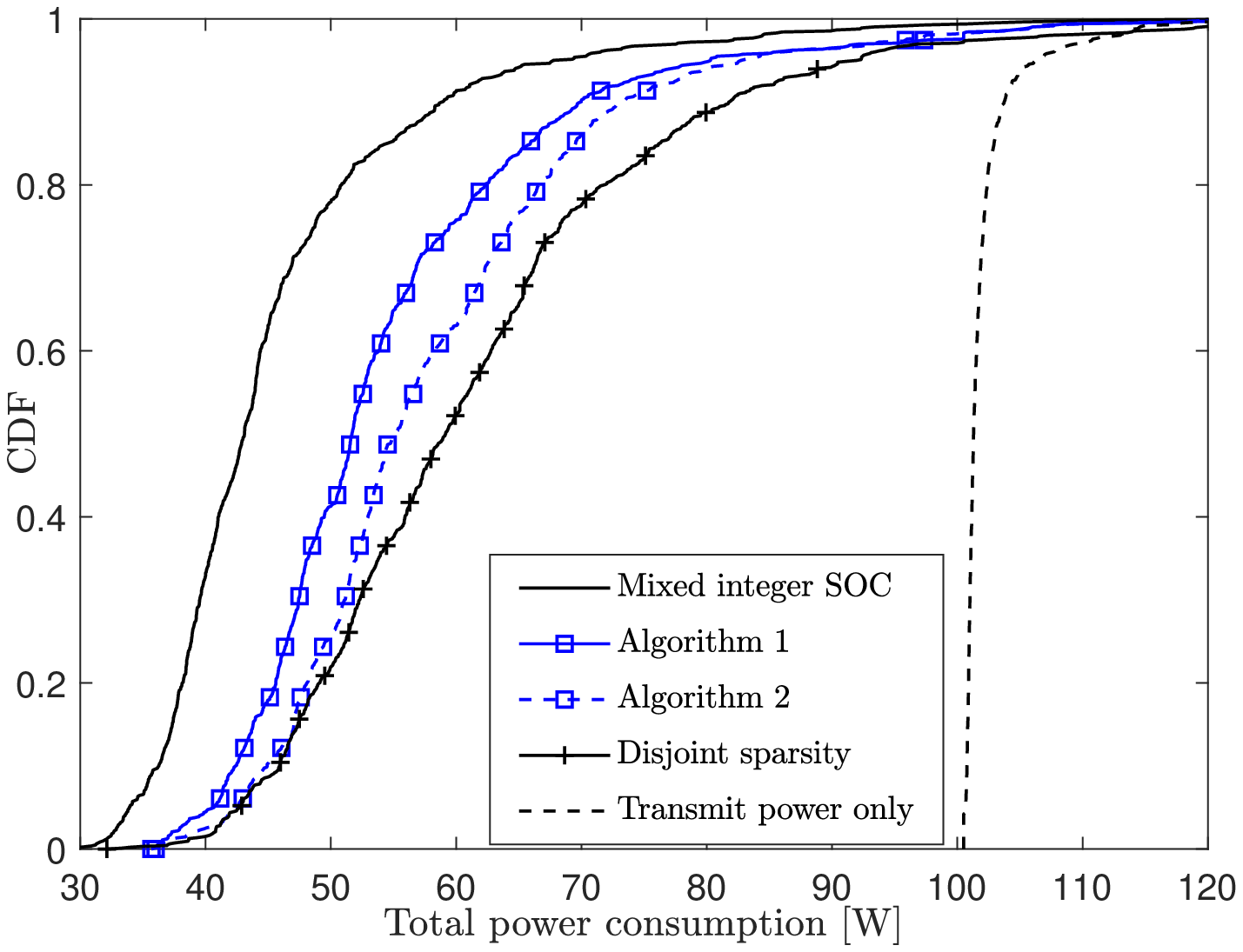} \vspace*{-0.25cm}
		\caption{The CDF of the total power consumption [W] with $M=20, K=20,$ and F-ZF precoding.}
		\label{FigTotalPowerZF}
		\vspace*{-0.4cm}
	\end{minipage}
\end{figure*}

\section{Numerical Results} \label{Sec:NumRes}
This section provides extensive numerical results to compare the power consumption of the different algorithms presented in the previous sections. We consider a Cell-free Massive MIMO network where the $M$ APs each having $N=20$ antennas and $K$ users are randomly distributed within a square of $1$~km$^2$. The distance between two APs should not be less than $50$~m. The requested SE of each user is $2$ b/s/Hz.\footnote{This value can be compared with the IMT-2020 requirement for 5G systems where the 5-th percentile SE is 0.225 b/s/Hz in dense urban scenarios. We demonstrate that one can achieve roughly 10 times more than that using Cell-free Massive MIMO.} We apply wrap-around to get rid of edge effects. %and guarantee uniform simulation performance for the $M$ APs.
The coherence intervals have $\tau_c = 200$ symbols. There are $\tau_p = 5$ orthogonal pilot signals and each is assigned to $K/\tau_p$ randomly selected users. We use the large-scale fading formulation in \cite{bjornson2019making}, that matches well with 3GPP Urban Microcell model with a carrier frequency $2$~GHz. In particular, the large-scale fading coefficient between user~$k$ and AP~$m$  is defined in dB-scale as
\begin{equation}
\beta_{mk} = -30.5 -36.7 \log_{10} (d_{mk} / 1 \textrm{m}) + z_{mk},
\end{equation}
where $d_{mk}$ is the distance which takes into account that APs are deployed 10\,m above the users. The shadow fading term $z_{mk}$ follows a Gaussian distribution with $z_{mk} \sim \mathcal{N}(0, 16)$. We assume the shadow fading coefficients from one AP to all users correlated as 
\begin{equation}
\mathbb{E} \{ z_{mk} z_{m'k'} \} = \begin{cases}
16 \times 2^{-\delta_{kk'}/ 9 \textrm{m}}, & \mbox{for } m = m', \\
0, & \mbox{for } m \neq m',
\end{cases}
\end{equation}
where $\delta_{kk'}$ is the distance between two users~$k$ and $k'$.

The power consumption model parameters are borrowed from \cite{ngo2018total}: The power amplifier inefficiency factor is $\Delta_m = 2.5$. The hardware power consumption per antenna is $0.2$~W and a fixed power consumption of each fronthaul link is setup to $0.825$~W, thus $P_m = 4.825$~W, $\forall m$. The traffic-dependent fronthaul power is $0.25$~W/(Gbits/s). The maximum transmit power budget per AP is $1$~W and pilot symbols have equal power $0.2$~W.

The following methods will be compared for either MRT or F-ZF precoding:\footnote{There are heuristic methods to turn off APs for energy-efficient purposes, but without any guarantee of \textbf{satisfying the SE requirements} \cite{Guillem2020}. There is no trivial benchmark that minimizes the total power consumption (both transmit and hardware powers) with respect to the SINR constraints.}
\begin{itemize}
\item[$(i)$] \textit{Total transmit power minimization only:} The network turns on all $M$ APs and optimizes the transmit powers for the given SE constraints. This case was considered in \cite{ngo2018total, nguyen2017energy}. We name this benchmark as \textit{Transmit power only} in the figures.
\item[$(ii)$] \textit{Algorithm~\ref{Algorithm:Sparsity}:} This algorithm first utilizes group-sparsity to order the APs and then selects how many APs to turnoff based on this ordering. We use $\tilde{p}/2 = 0.5$.
\item[$(iii)$] \textit{Algorithm~\ref{Algorithm:TotalTransPower}:} This algorithm uses the solution from ``Transmit power only'' to order the APs and then selects how many APs to turnoff based on this ordering.
\item[$(iv)$] \textit{Disjoint sparsity:} This is a method from  \cite{shi2016smoothed, luo2014downlink} that treats the selection of APs to turnoff and the transmit power minimization separately. We use $\tilde{p}/2 = 0.5$ and call this benchmark \textit{Disjoint sparsity} in the figures.
\item[$(v)$] \textit{Optimal solution:} This benchmark computes the optimal solution to both the transmit power allocation and active APs, as described in Lemma~\ref{Theorem:MixedInSOCP}. We name this benchmark as \textit{Mixed-integer SOC} in the figures.
\end{itemize}
\begin{figure*}[t]
	\begin{minipage}{0.48\textwidth}
		\centering
		\includegraphics[trim=0.5cm 0.0cm 1cm 0.5cm, clip=true, width=3.0in]{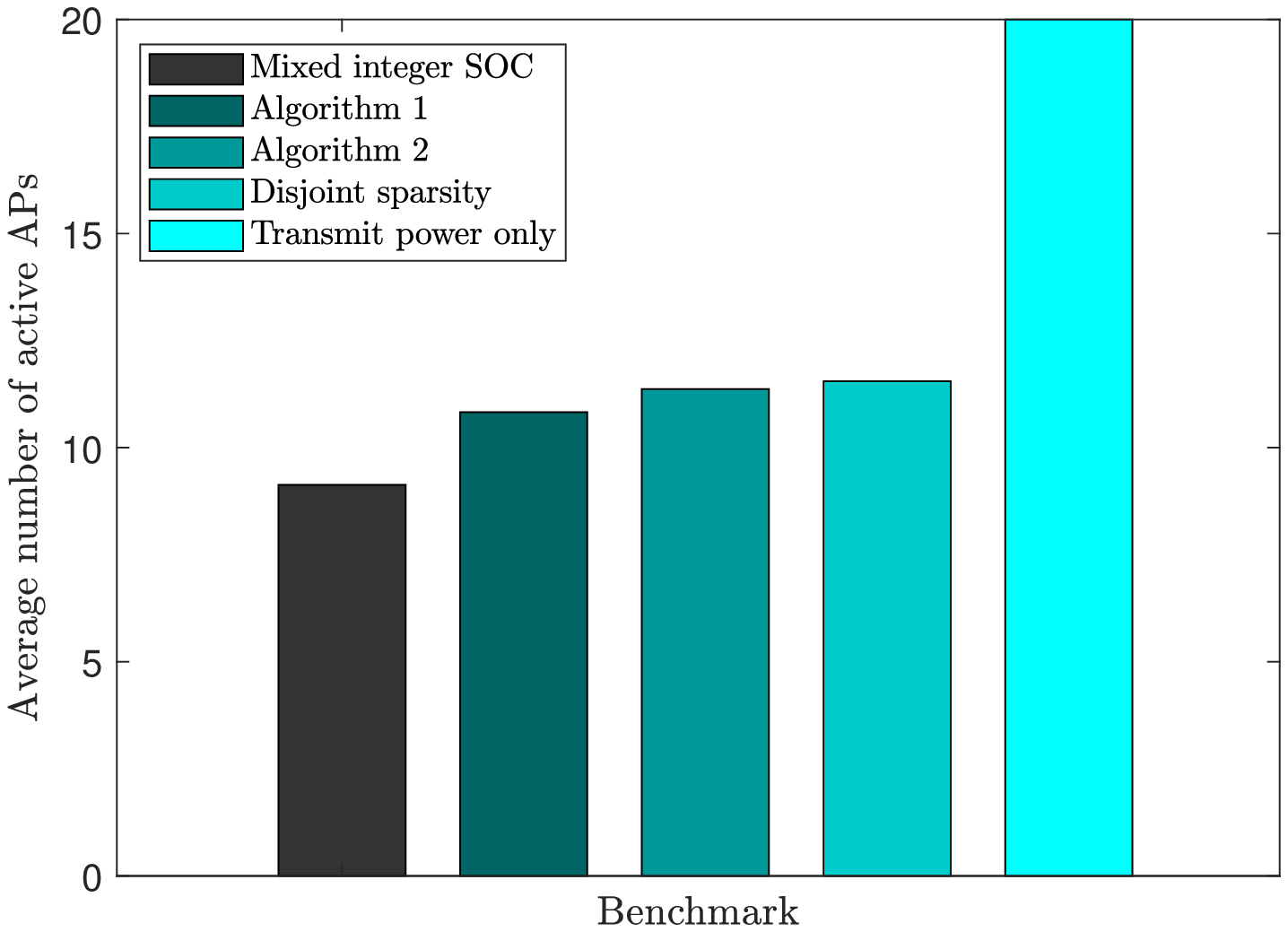} \vspace*{-0.25cm}
		\caption{The average number of active APs by using different benchmarks for a network with $M=20, K=20,$ and MRT precoding. }
		\label{FigActiveAPsMRT}
		\vspace*{-0.4cm}
	\end{minipage}
	\hfill
	\begin{minipage}{0.48\textwidth}
		\centering
		\includegraphics[trim=0.5cm 0.0cm 1cm 0.5cm, clip=true, width=3.0in]{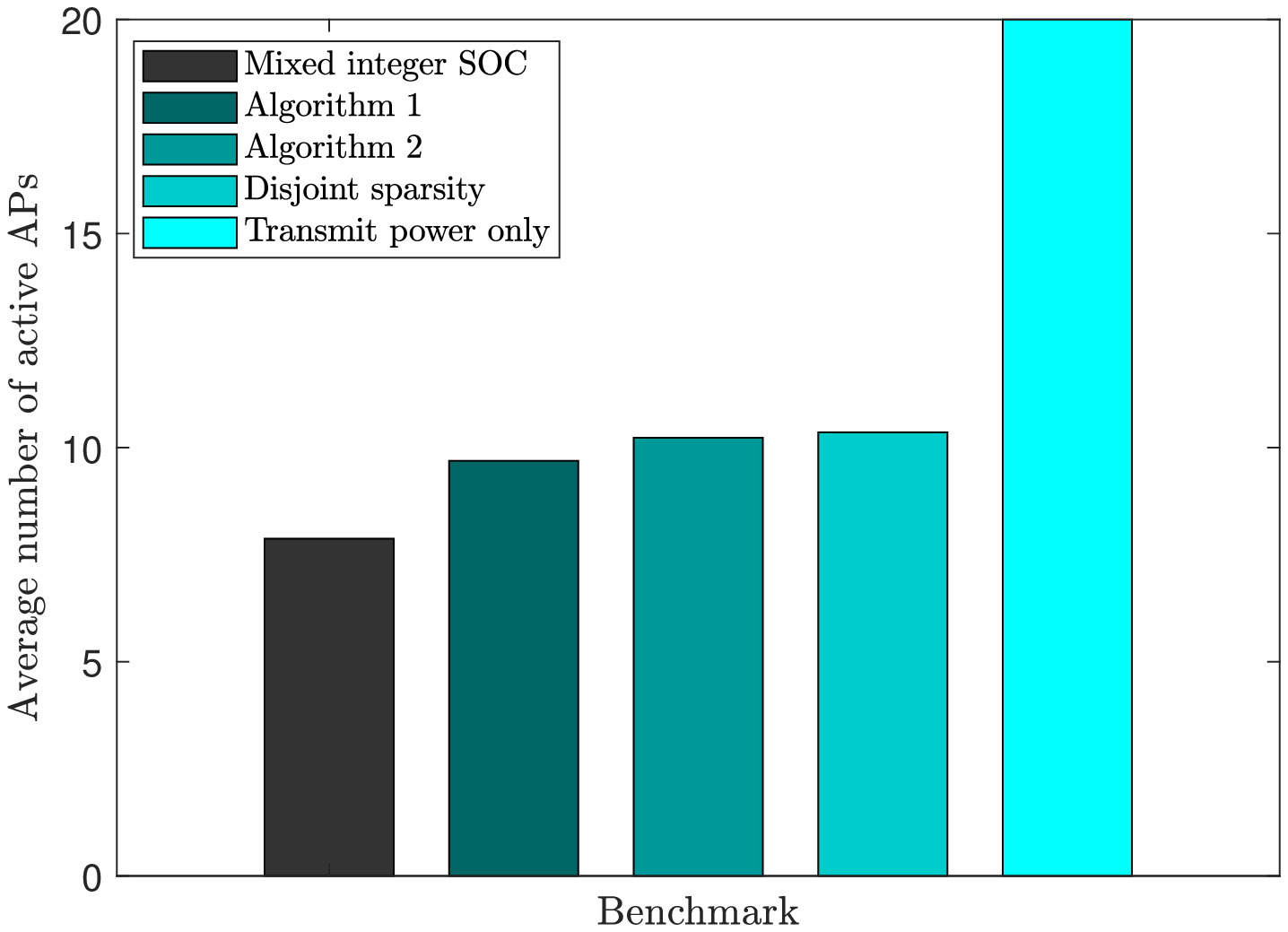} \vspace*{-0.25cm}
		\caption{The average number of active APs by using different benchmarks for a network with $M=20, K=20,$ and F-ZF precoding.}
		\label{FigActiveAPsZF}
		\vspace*{-0.4cm}
	\end{minipage}
\end{figure*}
\begin{figure}[t]
	\centering
	\includegraphics[trim=0.1cm 0.0cm 1cm 0.5cm, clip=true, width=3.0in]{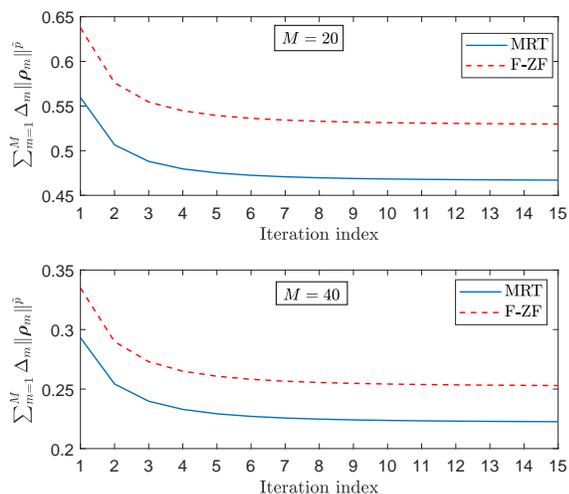} \vspace*{-0.5cm}
	\caption{The convergence of the group sparsity approach in Theorem~\ref{Theorem:Sparsity} for a network with the different number of APs serving $K=20$ users.}
	\label{FigConvergence}
	\vspace*{-0.4cm}
\end{figure}
Fig.~\ref{FigTransmitPowerMRT} shows the cumulative distribution function (CDF) of the total transmit power attained by the five different methods when using MRT precoding. When all APs are active, the average total transmit power is $1.8$~W. The mixed-integer SOC program needs an average transmit power of around $6.4$~W, which is $3.6 \times$ higher. Both the heuristic algorithms provide almost an equal average total transmit power of about $7$~W, while the disjoint sparsity benchmark uses the highest transmit power level: about $11.8$~W. At the $95\%$-likely point, the total transmit power is only $0.3$~W. Compared with this baseline, the mixed-integer SOC and the other methods require $8.5 \times$ and $4\times$ higher total transmit power.

We provide the CDFs of the total transmit power when using F-ZF precoding in Fig.~\ref{FigTransmitPowerZF}. F-ZF precoding reduces the total transmit power up to $12\%$ compared with MRT precoding. However, the minimum transmit power provided by F-ZF precoding is only $3\%$ lower than MRT precoding on the average. These gains by F-ZF precoding come from mitigating mutual interference as aforementioned in Sec.~\ref{Sec:DPL}. Moreover, the disjoint sparsity still consumes the highest transmit power with $10.2$~W on the average.
\begin{figure*}[t]
	\begin{minipage}{0.48\textwidth}
		\centering
		\includegraphics[trim=0.5cm 0.0cm 1cm 0.5cm, clip=true, width=3.0in]{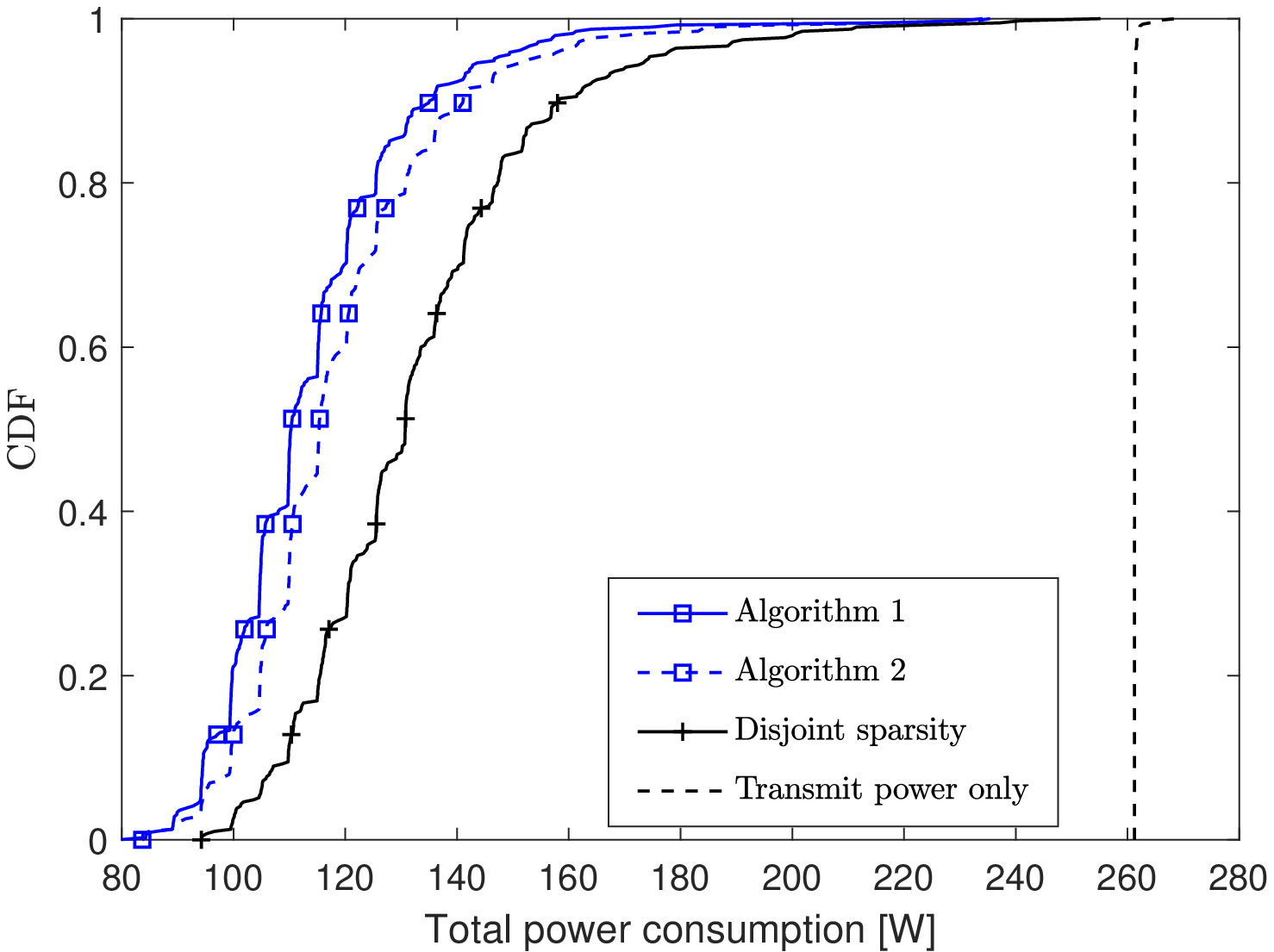} \vspace*{-0.25cm}
		\caption{The CDF of the total power consumption [W] with $M=50, K=40,$ and MRT precoding.}
		\label{FigTotalPowerK40M50MRT}
		\vspace*{-0.4cm}
	\end{minipage}
	\hfill
	\begin{minipage}{0.48\textwidth}
		\centering
		\includegraphics[trim=0.5cm 0.0cm 1cm 0.5cm, clip=true, width=3.0in]{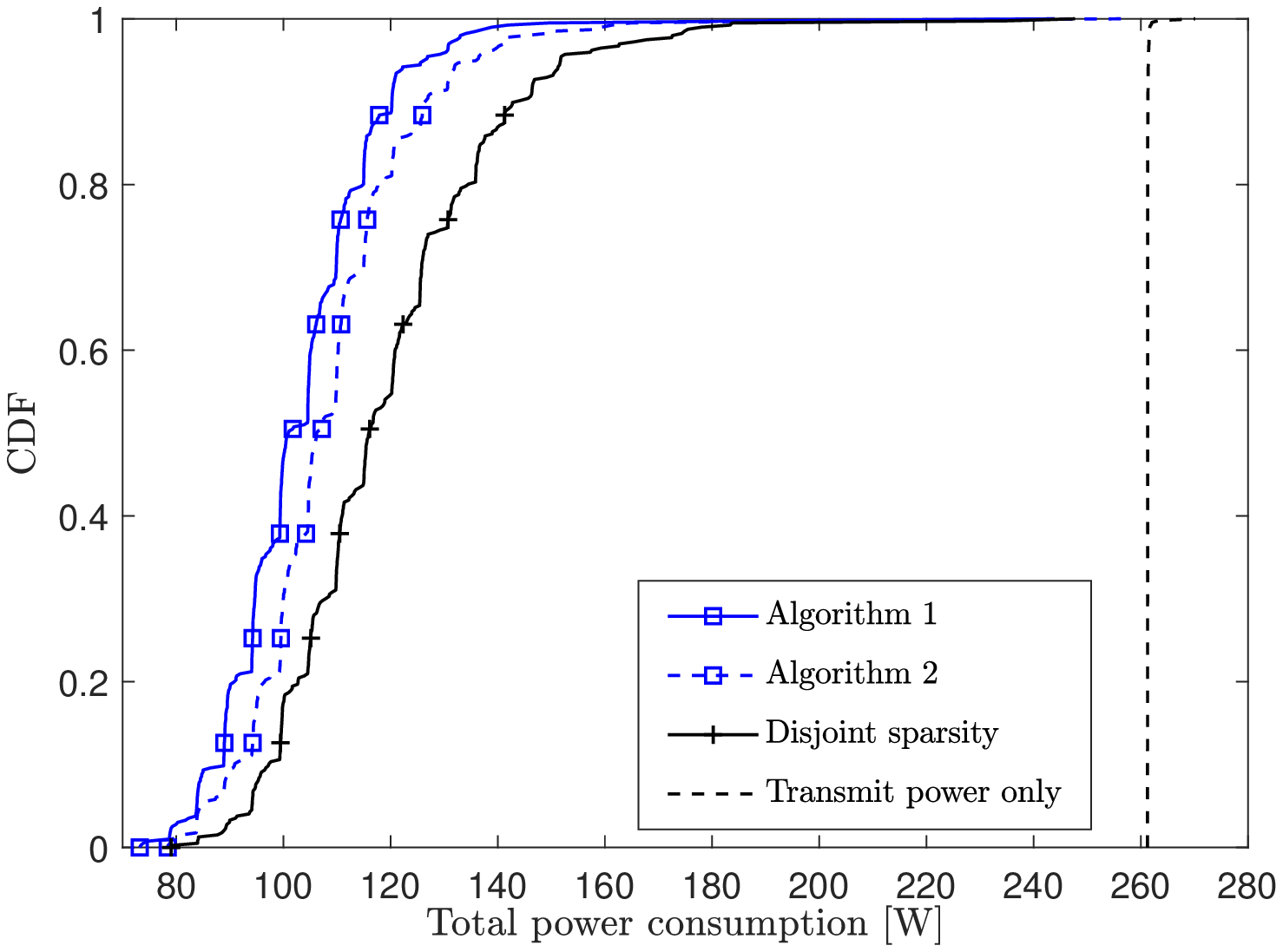} \vspace*{-0.25cm}
		\caption{The CDF of the total power consumption [W] with $M=50, K=40,$ and F-ZF precoding.}
		\label{FigTotalPowerK40M50ZF}
		\vspace*{-0.4cm}
	\end{minipage}
\end{figure*}

The proposed methods are not minimizing the transmit power but the total power consumption. Figs.~\ref{FigTotalPowerMRT} and \ref{FigTotalPowerZF} show the CDF of the total power consumption [W] of a network utilizing either MRT or F-ZF precoding. Contrary to the previous figures, ``Transmit power only'' now requires the highest total power consumption among all the considered methods for both MRT and F-ZF precoding: about $102$~W on the average. By solving the proposed mixed-integer SOC program, we find the global minimum total power consumption, which saves about $49\%$ or $55\%$ compared with the baseline by utilizing either MRT or F-ZF precoding, respectively. The proposed sparsity-based method in Algorithm~\ref{Algorithm:Sparsity} requires around $17\%$ and $20\%$ more power than the corresponding global minimum with MRT and F-ZF precoding, respectively. Since Algorithm~\ref{Algorithm:TotalTransPower} is not exploiting sparsity when selecting which APs to turn off, it requires $27\%$ extra power than the global minimum to serve all the users. Figs.~\ref{FigTotalPowerMRT} and \ref{FigTotalPowerZF} therefore confirm the improvements that can be made by exploiting the sparsity structure. Our proposed suboptimal algorithms give lower total power consumption than the disjoint sparsity benchmark with the improvement up to $59\%$ when MRT precoding is used. 

The average number of active APs  is plotted in Figs.~\ref{FigActiveAPsMRT} and \ref{FigActiveAPsZF}. The network only needs to activate a small subset of the APs to provide the required SEs. The mixed-integer SOC program requires around $9.1$ APs if MRT precoding is used, thus more than $55\%$ of the APs are in sleep mode. Algorithm~\ref{Algorithm:Sparsity} gives around $10.8$ active APs, while Algorithm~\ref{Algorithm:TotalTransPower} activates $11.3$ APs on average. The disjoint sparsity needs the number of active APs similar to Algorithm~\ref{Algorithm:TotalTransPower}, thus it proves the benefits of group sparsity in reducing the total transmit power consumption. If F-ZF precoding is applied, the network can turn off more APs compared with MRT, while still satisfying the SE requirements. For instance, the mixed-integer SOC program only needs $7.8$ active APs on the average.

Fig.~\ref{FigConvergence} shows the convergence of the proposed sparsity approach in Theorem~\ref{Theorem:Sparsity} for a network with $20$ or $40$ APs serving $20$ users, which is averaged over the $5000$ different realizations of user locations. The objective function on the vertical axis is defined as in \eqref{Prob:TotalPowerOptLpv1}. The results confirm the monotonically decreasing property, which was stated in Theorem~\ref{Theorem:Sparsity}. Compared with the initial point, the stationary point has a $20\%$ lower objective function value when the network has $20$ APs. The corresponding reduction is about $32\%$ when the network has $40$ APs. The proposed group sparsity approach converges to the stationary point in less than $15$ iterations, relatively irrespective of the number of APs. This shows an increasing cost of Algorithm~\ref{Algorithm:Sparsity} compared with Algorithm~\ref{Algorithm:TotalTransPower}.

Figs.~\ref{FigTotalPowerK40M50MRT} and \ref{FigTotalPowerK40M50ZF} show CDFs of the total power consumption for a network with $50$ APs serving $40$ users by utilizing MRT or F-ZF precoding, respectively. The mixed-integer SOC program is excluded in this case because of its extremely high complexity for large networks. 
``Transmit power only'' has the highest total power consumption with up to around $261$~W for both MRT or F-ZF precoding in use. Compared to this baseline, Algorithm~\ref{Algorithm:Sparsity} reduces the power consumption by $2.3\times$ if each AP uses MRT precoding and $2.5\times$ for the case of F-ZF precoding. Moreover, the group-sparsity structure provides at most $10\%$ lower power consumption than only deploying the optimized transmit powers as side information. We also confirm that jointly optimizing both transmit and hardware power gives better energy-efficiency than previous works, which treated the two classes of optimization variables in the disjoint sparsity approaches.
\begin{figure*}[t]
	\begin{minipage}{0.48\textwidth}
		\centering
		\includegraphics[trim=0.5cm 0.0cm 1cm 0.5cm, clip=true, width=3.0in]{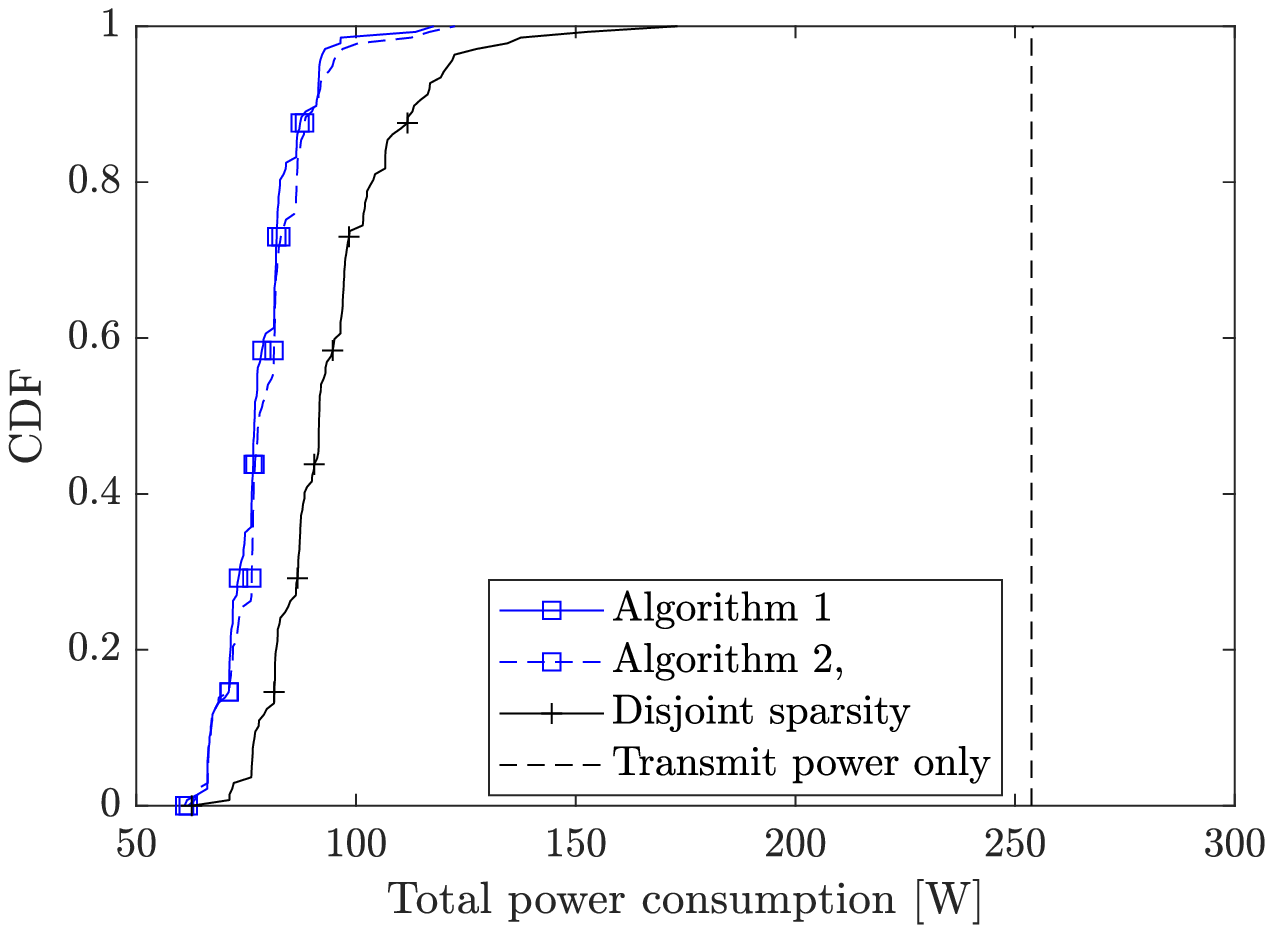} \vspace*{-0.25cm}
		\caption{The CDF of the total power consumption [W] with $M=50, K=40,$ and MRT precoding. Users has the different requested SEs.}
		\label{FigTotalPowerM50K40MRTVariousQoS}
		\vspace*{-0.4cm}
	\end{minipage}
	\hfill
	\begin{minipage}{0.48\textwidth}
		\centering
		\includegraphics[trim=0.5cm 0.0cm 1cm 0.5cm, clip=true, width=3.0in]{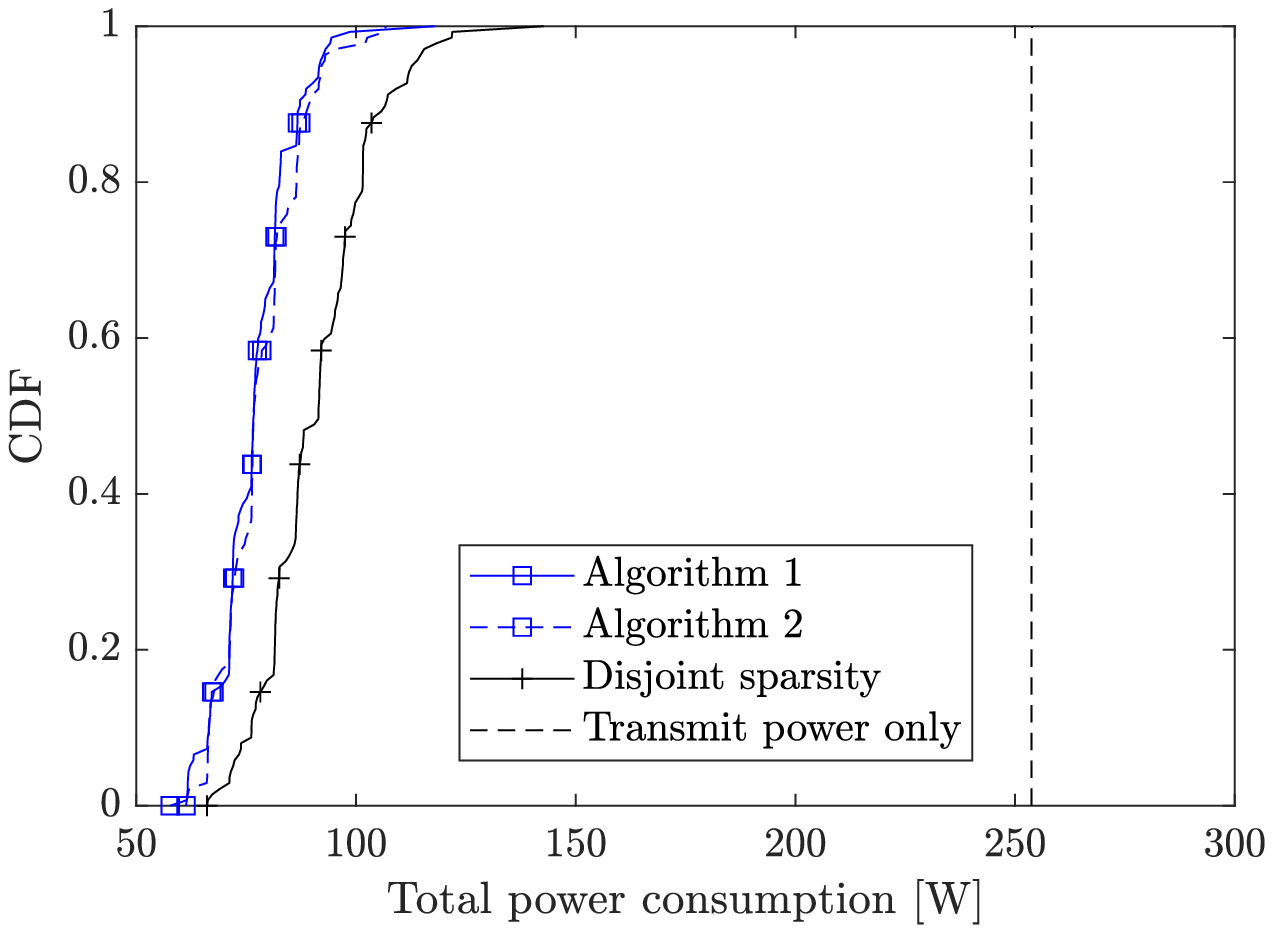} \vspace*{-0.25cm}
		\caption{The CDF of the total power consumption [W] with $M=50, K=40,$ and F-ZF precoding. Users has the different requested SEs.}
		\label{FigTotalPowerM50K40ZFVariousQoS}
		\vspace*{-0.4cm}
	\end{minipage}
\end{figure*}

Figs.~\ref{FigTotalPowerM50K40MRTVariousQoS} and \ref{FigTotalPowerM50K40ZFVariousQoS} consider a similar setup as in the two previous figures, but each user asks for a different SE that is drawn from a uniform distribution between 1 and 2 b/s/Hz. 
The case where only the transmit power is optimized is still requiring the highest power, followed by the disjoint sparsity method.
Algorithm~\ref{Algorithm:Sparsity} requires a $3\times$ lower average total power consumption than when only the transmit power is minimized.
Moreover, the effectiveness of Algorithm~\ref{Algorithm:TotalTransPower} is observed from the fact that it substantially reduces the complexity compared to Algorithm~\ref{Algorithm:Sparsity}, but only requires $2\%$ more power.
Besides, Figs.~\ref{FigTotalPowerM50K40MRTVariousQoS} and \ref{FigTotalPowerM50K40ZFVariousQoS} show that the total power consumption is always less than those of Figs.~\ref{FigTotalPowerK40M50MRT} and \ref{FigTotalPowerK40M50ZF} when all users request the same SE of $2$ b/s/Hz.

\section{Conclusion} \label{Sec:Conclusion}
This paper has minimized the total power consumption optimization in Cell-free Massive MIMO networks by jointly optimizing the downlink transmit powers and the number of active APs, while satisfying the SEs requested by all the users. A globally optimal solution can be found by formulating the considered problem as a mixed-integer SOC program then utilizing the branch-and-bound approach. From this joint optimization framework, we observe a considerable reduction of the total power consumption (up to more than $50\%$) compared with only minimizing the transmit power as in previous work. Due to the high computational complexity of solving the mixed-integer SOC program, we developed two suboptimal algorithms that have a complexity that make them applicable also in large cell-free networks. These algorithms only require roughly $20\%$ higher power consumption than the global optimum. %For the future works, we can consider the design of more efficient suboptimal algorithms with even lower computational complexity and closer to the global optimum. 

\appendix
%\subsection{Useful Lemma}
%\begin{lemma}[H\"{o}lder inequality \cite{beckenbach2012inequalities}]
%For $x_i, z_i \geq 0, \forall i= 1, \ldots, n$ and $r>1, s >1$ satisfying $r^{-1} + s^{-1} = 1$, it holds that
%\begin{equation}
%\sum_{i=1}^n x_i z_i \leq \left( \sum_{i=1}^n x_i^r \right)^{-r} \left( %\sum_{i=1}^n z_i^s \right)^{-s},
%\end{equation} 
%where the equality happens if and only if $x_i^{r} z_i^{-s} = \ldots = x_{i'}^{r} z_{i'}^{-s}$ for all positive values $x_i, x_{i'}, z_{i}, z_{i'}$.
%\end{lemma}
%\subsection{Solving Problem~\eqref{Prob:BnB} by the Branch and Bound} \label{Appendix:BnB}
\subsection{Proof of Lemma~\ref{Lemma:ComplexityBnB}} \label{Appendix:ComplexityBnB}
The proof is based on formulating and computing the computational complexity of every SOC program for the use in the branch-and-bound approach. Iteration~$n$ selects a box $\mathcal{B}^{(n)}$ with the $M_0^{(n-1)}$ zeros, the $M_1^{(n-1)}$ ones, and the $Q^{(n)}$ unfixed binary variables. The new two boxes $\mathcal{B}_0^{(n)}$ and $\mathcal{B}_1^{(n)}$ are further generated from $\mathcal{B}^{(n)}$ by fixing one of the $Q^{(n)}$ binary variables, say $\alpha_{m'}$, as:
\begin{equation}
\mathcal{B}_i^{(n)} = \left\{ \mathcal{B}^{(n)} | \alpha_{m'} = i, i \in \{0,1\} \right\}.
\end{equation}
Let us denote $\mathcal{B}_{i1}^{(n)} \subseteq \mathcal{B}_i^{(n)}$ the subset containing all ``known" active APs and $\mathcal{B}_{iu}^{(n)} \subseteq \mathcal{B}_i^{(n)}$ the subset containing the unfixed APs (i.e., $\mathcal{B}_{i1}^{(n)} \cup \mathcal{B}_{iu}^{(n)} =  \mathcal{B}_i^{(n)}$). Based on the new box $\mathcal{B}_i^{(n)}$, a lower bound on the global optimum is computed by solving the following SOC program:
\begin{equation} \label{Prob:BnBv1} 
\begin{aligned}
\mathsf{lb}(\mathcal{B}_i^{(n)}):	& \underset{ \{ \rho_{mk} \geq 0 \}, \{ \alpha_m \}, s }{\mathrm{minimize}}
& &  s \\
& {\mathrm{subject \,\, to}}
&&  \| \mathbf{r} \| \leq  s,   \\
& & & \| \mathbf{s}_{k} \| \leq \mathbf{g}_{k}^T \mathbf{u}_{k}, \; \forall k = 1, \ldots, K,  \\
& & & \| \tilde{\mathbf{u}}_{m}'\| \leq  \sqrt{P_{\mathrm{max},m}}, \; \forall m \in \mathcal{B}_{i1}^{(n)}, \\
& & & \| \tilde{\mathbf{u}}_{m}'\| \leq \alpha_m \sqrt{P_{\mathrm{max},m}}, \; \forall m \in \mathcal{B}_{iu}^{(n)}, \\
&&& 0 \leq \alpha_m \leq 1, \forall m \in \mathcal{B}_{iu}^{(n)}.
\end{aligned}
\end{equation}
The SOC program \eqref{Prob:BnBv1} includes $L_{i2}^{(n)}$ optimization variables, one SOC constraint of dimension $L_{i2}^{(n)}$, the $K$ SOC constraints of dimension with the $k$-th one being of dimension $\tilde{K}_k$, the $L_{i1}^{(n)}$ constraints of dimension $K$, and the $Q^{(n)} -1$ constraints of dimension $K+1$. By following similar steps as in \cite{van2018distributed, wang2014outage}, we can compute the per-iteration computation costs to solve an SOC program and the order of the number of iterations to reach $\varepsilon$-accuracy as
\begin{align}
& \mathcal{O} \left( \left(L_{i2}^{(n)} \right)^3 +      L_{i2}^{(n)}\sum_{k=1}^K \left| \mathcal{P}_k \right| + L_{i2}^{(n)} L_{i1}^{(n)} K^2 + Z^{(n)} K \right), \label{eq:lbB0nv1}\\
%\end{align}
%\begin{align}
& \sqrt{L_{i2}^{(n)}  + K^2  + L_{i1}^{(n)}K + Z^{(n)}} \ln \big( \varepsilon^{-1} \big)  \label{eq:lbB0nv2}.
\end{align}
Taking the product of \eqref{eq:lbB0nv1} and \eqref{eq:lbB0nv2}, the computational complexity to obtain $\mathsf{lb}(\mathcal{B}_i^{(n)})$ is $\ln \big( \varepsilon^{-1} \big)\mathcal{O} \left( C_i^{(n),\mathrm{lb}} \right)$. 

If we denote $\alpha_m^{\ast}$ the optimal solution to $\alpha_m$ in problem~\eqref{Prob:BnBv1}, then $\tilde{\alpha}_m \in \{0, 1\}, \forall m,$ will be the binary number obtained by using the rounding operator to each element in the set $\{\alpha_m^{\ast} \}$. An upper bound based on a new box $\mathcal{B}_i^{(n)}$ is obtained by solving this SOC program:
\begin{equation} \label{Prob:BnBv2} 
\begin{aligned}
\mathsf{ub}(\mathcal{B}_i^{(n)}): \, & \underset{ \{ \rho_{mk} \geq 0 \},  s }{\mathrm{minimize}}
& &  s \\
& {\mathrm{subject \,\, to}}
&&  \| \mathbf{r} \| \leq  s,   \\
& & & \| \mathbf{s}_{k} \| \leq \mathbf{g}_{k}^T \mathbf{u}_{k}, \; \forall k = 1, \ldots, K,  \\
& & & \| \tilde{\mathbf{u}}_{m}'\| \leq  \sqrt{P_{\mathrm{max},m}}, \; \forall m \in \mathcal{B}_{i1}^{(n)}, \\
& & & \| \tilde{\mathbf{u}}_{m}'\| \leq \tilde{\alpha}_m \sqrt{P_{\mathrm{max},m}}, \; \forall m \in \mathcal{B}_{iu}^{(n)}.
\end{aligned}
\end{equation}
The SOC program \eqref{Prob:BnBv2} has $U_{i}^{(n)}K+1$ optimization variables, one SOC constraint of dimension $U_{i}^{(n)}K+1$, the $K$ SOC constraints with the $k$-th one of dimension $\tilde{K}_k$, and the $U_{i}^{(n)}$ constraints of dimension $K$. The  computation cost to reach $\varepsilon$-accuracy is computed as
\begin{align}
& \mathcal{O} \left( \big( U_i^{(n)} \big)^3 K^3  \sqrt{U_i^{(n)}K+ K^2 } \right)\ln \big( \varepsilon^{-1} \big), \label{eq:ubB0nv2} 
\end{align} 
which is the computational complexity order to obtain $\mathsf{ub}(\mathcal{B}_i^{(n)})$ as denoted by $\ln \big( \varepsilon^{-1} \big) C_i^{(n),\mathrm{ub}}$ in the lemma. The computational complexity order of the branch-and-bound approach is obtained by summing up the costs over $N_1$ iterations.
\subsection{Proof of Theorem~\ref{Theorem:Sparsity}} \label{Appendix:Sparisity}
We first prove that the iterative process in Theorem~\ref{Theorem:Sparsity} produces a non-increasing objective function of problem~\eqref{Prob:TotalPowerOptLpv1}, mathematically expressed as
\begin{equation}
\sum_{ m=1}^M \Delta_m \| \pmb{\rho}_{m}^{\ast,(n)} \|^{\tilde{p}} \leq \sum_{ m=1}^M \Delta_m \| \pmb{\rho}_{m}^{\ast,(n-1)} \|^{\tilde{p}}. 
\end{equation}
Indeed, the following series of inequalities holds
\begin{equation}
\begin{split}
&\sum_{m=1}^M  \frac{\Delta_m \tilde{p}}{2} \| \pmb{\rho}_{m}^{\ast,(n)} \|^{\tilde{p}}  \stackrel{(a)}{=} \underset{\pmb{\rho}_{m}^{(n)} \succeq \mathbf{0} }{\min} \sum_{m=1}^M a_m^{(n-1)}  \| \pmb{\rho}_{m}^{(n)} \|^2   \\
& \stackrel{(b)}{=}  \underset{\pmb{\rho}_{m}^{(n)} \succeq \mathbf{0} }{\min} \sum_{m=1}^M \frac{\Delta_m \tilde{p}}{2} \left(\| \pmb{\rho}_{m}^{\ast,(n-1)} \|^2 + \epsilon_{n-1}^2 \right)^{\frac{\tilde{p}}{2} -1}  \| \pmb{\rho}_{m}^{(n)} \|^2\\
& \stackrel{(c)}{\leq} \sum_{m=1}^M \frac{\Delta_m \tilde{p}}{2} \left( \| \pmb{\rho}_{m}^{\ast,(n-1)} \|^2 + \epsilon_{n-1}^2 \right)^{\frac{\tilde{p}}{2} -1}  \| \pmb{\rho}_{m}^{\ast,(n-1)} \|^2  \\
&\stackrel{(d)}{\leq} \sum_{m=1}^M  \frac{\Delta_m \tilde{p}}{2} \| \pmb{\rho}_{m}^{\ast,(n-1)} \|^{\tilde{p}},
\end{split}
\end{equation}
where $(a)$ and $(c)$ are obtained from solving problem~\eqref{Prob:TotalPowerOptv3}, while $(b)$ and $(d)$ are due to the definition of the weight values in \eqref{eq:weightv1}. Therefore, the iteratively weighted least squares approach, which is applied in this paper, converges to a fixed point. Let  $f = \sum_{ m=1}^M \Delta_m \| \pmb{\rho}_{m} \|^{\tilde{p}}$ denote the objective function of problem~\eqref{Prob:TotalPowerOptLpv1}, then taking the first derivative of $f$ with respect to $ \rho_{mk} = \rho_{mk}^{\ast, (n)} $ yields
\begin{equation} \label{eq:FirstDev}
\begin{split}
\frac{\partial f}{\partial \sqrt{\rho_{mk}} } \Big|_{\rho_{mk} = \rho_{mk}^{\ast, (n)}} &= \frac{\Delta_m \tilde{p}}{2} \| \pmb{\rho}_{m}^{\ast,(n)} \|^{2 \left( \frac{\tilde{p}}{2} -1\right)} 2 \sqrt{ \rho_{mk}^{\ast, (n)}} \\
& \stackrel{(a)}{\rightarrow} a_{m}^{(n)} 2 \sqrt{ \rho_{mk}^{\ast, (n)}},
\end{split}
\end{equation}
where $(a)$ is obtained with a sufficiently large number of iterations such that $\epsilon_n \rightarrow 0 $. According to \cite[Proposition 2.1.2]{bertsekas1999nonlinear}, if the fixed point holds at iteration~$n$, then the following optimality condition is obtained
\begin{equation} \label{eq:ConditionFixedP}
\sum_{m = 1}^M \sum_{k=1}^K a_{m}^{(n)}  \sqrt{ \rho_{mk}^{\ast, (n)}} \left(  \sqrt{ \rho_{mk}} -  \sqrt{ \rho_{mk}^{\ast, (n)}} \right) \geq 0.
\end{equation}
Substituting \eqref{eq:FirstDev} into \eqref{eq:ConditionFixedP}, the local minimum of $f$ is established as
\begin{equation}
\sum_{m = 1}^M \sum_{k=1}^K \frac{\partial f}{\partial \sqrt{\rho_{mk}} } \Big|_{\rho_{mk} = \rho_{mk}^{\ast, (n)}}  \left(  \sqrt{ \rho_{mk}} -  \sqrt{ \rho_{mk}^{\ast, (n)}} \right) \geq 0,
\end{equation}
which confirms that $\{ \rho_{mk}^{\ast, (n)} \}$ is a stationary point to problem~\eqref{Prob:TotalPowerOptLpv1} by virtue of \cite[Definition~$1$]{ba2014}. To prove the second property, we observe the following inequality
\begin{equation} \label{eq:ASSv1}
\begin{split}
&\sum_{m=1}^M a_m^{(n-1)} \| \pmb{\rho}_{m}^{\ast,(n)} \|^2 \geq a_{m'}^{(n-1)} \| \pmb{\rho}_{m'}^{\ast,(n)} \|^2 \\
&= \frac{\Delta_{m'} \tilde{p} }{2} \left( \| \pmb{\rho}_{m'}^{\ast,(n-1)} \|^2 + \epsilon_{n-1}^2 \right)^{\frac{\tilde{p}}{2} -1}  \| \pmb{\rho}_{m'}^{\ast,(n)} \|^2,
\end{split}
\end{equation}
with $m' \in \{ 1, \ldots, M \}$. Moreover, the feasibility domain of problem~\eqref{Prob:TotalPowerOptv3} at iteration~$n$ gives the relationship 
\begin{equation} \label{eq:ASS2}
\sum_{m=1}^M a_m^{(n-1)} \| \pmb{\rho}_{m}^{\ast,(n)} \|^2 \leq  \sum_{m=1}^M \frac{\Delta_m \tilde{p}}{2}  \| \pmb{\rho}_{m}^{\ast,(n-1)} \|^{\tilde{p}}.
\end{equation}
Combining \eqref{eq:ASSv1} and \eqref{eq:ASS2}, we obtain the following inequality:
\begin{align}
%\begin{split}
& \Delta_{m'}  \left( \| \pmb{\rho}_{m'}^{\ast,(n-1)} \|^2 + \varepsilon_{n-1}^2 \right)^{\frac{\tilde{p}}{2} -1}  \| \pmb{\rho}_{m'}^{\ast,(n)} \|^2  \leq \sum_{m=1}^M \Delta_m   \| \pmb{\rho}_{m}^{\ast,(n-1)} \|^{\tilde{p}}, \\
 \Leftrightarrow & \Delta_{m'}  \| \pmb{\rho}_{m'}^{\ast,(n)} \|^2  \leq \left(\| \pmb{\rho}_{m'}^{\ast,(n-1)} \|^2 + \epsilon_{n-1}^2 \right)^{ 1 - \frac{\tilde{p}}{2}  }  \sum_{m=1}^M \Delta_m  \| \pmb{\rho}_{m}^{\ast,(n-1)} \|^{\tilde{p}}. \label{eq:BoundOneVal}
%\end{split}
\end{align}
When $\epsilon_{n-1} \rightarrow 0$ as $n \rightarrow \infty$, \eqref{eq:BoundOneVal} is approximated as
\begin{equation}
\Delta_{m'}  \| \pmb{\rho}_{m'}^{\ast,(n)} \|^2  \leq \| \pmb{\rho}_{m'}^{\ast,(n-1)} \|^{ 2\left(1 - \frac{\tilde{p}}{2}\right)  }  \sum_{m=1}^M \Delta_m  \| \pmb{\rho}_{m}^{\ast,(n-1)} \|^{\tilde{p}},
\end{equation}
which indicates that if iteration $n-1$ gives the total transmit power of AP~$m'$ equal to zero, i.e.,  $\| \pmb{\rho}_{m'}^{\ast,(n-1)} \| =0$, then 
\begin{equation} \label{eq:Sparsityv1}
\| \pmb{\rho}_{m'}^{\ast,(n)} \|^2 \leq 0.
\end{equation}
From~\eqref{eq:Sparsityv1}, it indicates at iteration~$n$ that $\{\rho_{m'k}^{\ast,(n)} = 0, \forall k= 1,\ldots, K \}$. 

%======================== ==Reference==========================================================================
\bibliographystyle{IEEEtran}
\bibliography{IEEEabrv,refs}
\begin{IEEEbiography} 
	%[{\includegraphics[width=1.0in,height=1.25in,clip,keepaspectratio]{images/TrinhVanChien.jpg}}]
	{Trinh Van Chien} received the B.S. degree in Electronics and Telecommunications from Hanoi University of Science and Technology (HUST), Vietnam, in 2012. He then received the M.S. degree in Electrical and Computer Enginneering from Sungkyunkwan University (SKKU), Korea, in 2014 and the Ph.D. degree in Communication Systems from Link\"oping University (LiU), Sweden, in 2020. His interest lies in convex optimization problems for wireless communications and image \& video processing. He was an IEEE wireless communications letters exemplary reviewer for 2016 and 2017. He also received the award of scientific excellence in the first year of the 5Gwireless project funded by European Union Horizon's 2020. 
\end{IEEEbiography}
\begin{IEEEbiography} 
	%[{\includegraphics[width=1.0in,height=1.25in,clip,keepaspectratio]{images/EmilBjornson.jpg}}]
	{Emil Bj\"ornson} (S'07-M'12-SM'17) received the M.S. degree in engineering mathematics from Lund University, Sweden, in 2007, and the Ph.D. degree in telecommunications from the KTH Royal Institute of Technology, Sweden, in 2011. From 2012 to 2014, he held a joint post-doctoral position at the Alcatel-Lucent Chair on Flexible Radio, SUPELEC, France, and the KTH Royal Institute of Technology. He joined Link\"oping University, Sweden, in 2014, where he is currently an Associate Professor and a Docent with the Division of Communication Systems.
	
	He has authored the textbooks \emph{Optimal Resource Allocation in Coordinated Multi-Cell Systems} (2013) and \emph{Massive MIMO Networks: Spectral, Energy, and Hardware Efficiency} (2017). He is dedicated to reproducible research and has made a large amount of simulation code publicly available. He performs research on MIMO communications, radio resource allocation, machine learning for communications, and energy efficiency. Since 2017, he has been on the Editorial Board of the IEEE TRANSACTIONS ON COMMUNICATIONS and the IEEE TRANSACTIONS ON GREEN COMMUNICATIONS AND NETWORKING since 2016.
	
	He has performed MIMO research for over ten years and has filed more than twenty MIMO related patent applications. He has received the 2014 Outstanding Young Researcher Award from IEEE ComSoc EMEA, the 2015 Ingvar Carlsson Award, the 2016 Best Ph.D. Award from EURASIP, the 2018 IEEE Marconi Prize Paper Award in Wireless Communications, the 2019 EURASIP Early Career Award, the 2019 IEEE Communications Society Fred W. Ellersick Prize, and the 2019 IEEE Signal Processing Magazine Best Column Award. He also co-authored papers that received Best Paper Awards at the conferences, including WCSP 2009, the IEEE CAMSAP 2011, the IEEE WCNC 2014, the IEEE ICC 2015, WCSP 2017, and the IEEE SAM 2014.  
\end{IEEEbiography}
\begin{IEEEbiography} 
	%[{\includegraphics[width=1.0in,height=1.25in,clip,keepaspectratio]{images/ErikGLarsson.jpg}}]
	{Erik G. Larsson} (S'99--M'03--SM'10--F'16)
	received the Ph.D. degree from Uppsala University,
	Uppsala, Sweden, in 2002.  He is currently Professor of Communication
	Systems at Link\"oping University (LiU) in Link\"oping, Sweden. He was
	with the KTH Royal Institute of Technology in Stockholm, Sweden, the
	George Washington University, USA, the University of Florida, USA, and
	Ericsson Research, Sweden.  His main professional interests are within
	the areas of wireless communications and signal processing. He 
	co-authored \emph{Space-Time Block Coding for  Wireless Communications} (Cambridge University Press, 2003) 
	and \emph{Fundamentals of Massive MIMO} (Cambridge University Press, 2016). 
	He is co-inventor of 19 issued U.S. patents.
	
	Currently he is an editorial board member of the \emph{IEEE Signal
		Processing Magazine}, and a member of the  \emph{IEEE Transactions on Wireless Communications}    steering committee. 
	He served as  chair  of the IEEE Signal Processing Society SPCOM technical committee (2015--2016), 
	chair of  the \emph{IEEE Wireless  Communications Letters} steering committee (2014--2015), 
	General respectively Technical Chair of the Asilomar SSC conference (2015, 2012), 
	technical co-chair of the IEEE Communication Theory Workshop (2019), 
	and   member of the  IEEE Signal Processing Society Awards Board (2017--2019).
	He was Associate Editor for, among others, the \emph{IEEE Transactions on Communications} (2010-2014) 
	and the \emph{IEEE Transactions on Signal Processing} (2006-2010).
	
	He received the IEEE Signal Processing Magazine Best Column Award
	twice, in 2012 and 2014, the IEEE ComSoc Stephen O. Rice Prize in
	Communications Theory in 2015, the IEEE ComSoc Leonard G. Abraham
	Prize in 2017, the IEEE ComSoc Best Tutorial Paper Award in 2018, and
	the IEEE ComSoc Fred W. Ellersick Prize in 2019.  
\end{IEEEbiography}
\end{document}